\documentclass[aps,physrev,reprint,amsmath,amssymb,floatfix]{revtex4-2}

\usepackage{anyfontsize}
\usepackage{amsmath,amssymb,amsthm,mathtools}
\usepackage{booktabs,tabularx,array,multirow}
\usepackage{graphicx}
\graphicspath{{./figures/}}
\usepackage{xcolor}
\usepackage{microtype}
\usepackage{enumitem}
\usepackage[caption=false]{subfig}
\usepackage{hyperref}
\usepackage[nameinlink]{cleveref}
\crefname{figure}{Fig.}{Figs.}
\Crefname{figure}{Fig.}{Figs.}
\crefname{equation}{Eq.}{Eqs.}
\Crefname{equation}{Eq.}{Eqs.}
\crefname{section}{Sec.}{Secs.}
\Crefname{section}{Sec.}{Secs.}
\crefname{subsection}{Sec.}{Secs.}
\Crefname{subsection}{Sec.}{Secs.}
\crefname{table}{Table}{Tables}
\Crefname{table}{Table}{Tables}
\hypersetup{
  hidelinks,
  pdftitle={Requirements for Synthetic-Frequency Switching in Integrated Photonic Fabrics},
  pdfauthor={Jorge Parra},
  pdfsubject={Architectural roles and device requirements of synthetic-frequency coupling in photonic switching fabrics},
  pdfkeywords={photonic switching, synthetic-frequency photonics, frequency-mode coupling, thin-film lithium niobate, optical circuit switching}
}

\definecolor{spatial}{HTML}{2563EB}
\definecolor{modal}{HTML}{D97706}
\definecolor{electrical}{HTML}{64748B}
\definecolor{synthetic}{HTML}{059669}

\IfFileExists{results/generated_macros.tex}{

\newcommand{\GlobalBaseline}{60.9\%}
\newcommand{\GlobalRFOne}{39.5\%}
\newcommand{\GlobalRFThree}{29.9\%}
\newcommand{\GlobalFull}{28.1\%}

}{

  \newcommand{\GlobalBaseline}{--}
  \newcommand{\GlobalRFOne}{--}
  \newcommand{\GlobalRFThree}{--}
  \newcommand{\GlobalFull}{--}

}
\IfFileExists{results/revision_macros.tex}{

\newcommand{\RevisionFRecAThreeQEight}{94.7\%}

\newcommand{\RevisionFRecAThreeQSixteen}{84.2\%}

\newcommand{\RevisionComplexityFRecAThreeQFour}{100.0\%}

\newcommand{\RevisionComplexityFRecAThreeQEight}{95.2\%}
\newcommand{\RevisionComplexityFRecAThreeQEightCI}{2.3}

\newcommand{\RevisionComplexityFRecAThreeQSixteen}{91.0\%}

\newcommand{\RevisionFixedCostBOneTwoBlocking}{32.9\%}

\newcommand{\RevisionFixedCostBOneThreeBlocking}{32.1\%}

\newcommand{\RevisionFixedCostBOneThreeUtilizationOne}{25.7\%}

\newcommand{\RevisionFixedCostBOneThreeUtilizationThree}{21.4\%}

\newcommand{\RevisionFixedCostBTwoThreeBlocking}{32.9\%}

}{}
\IfFileExists{results/tfln_physical_robustness_macros.tex}{

}{}
\IfFileExists{results/tfln_nested_reference_macros.tex}{
\newcommand{\NestedDOneTwoBlocking}{60.4\%}

\newcommand{\NestedRFOffBlocking}{54.3\%}

\newcommand{\NestedBypassBoundBlocking}{42.1\%}

\newcommand{\NestedRichJThree}{0.004}

\newcommand{\NestedDOneTwoLossOneFSR}{6.06}
\newcommand{\NestedDOneTwoLossTwoFSR}{10.76}
\newcommand{\NestedDOneTwoIsolationOneFSR}{3.18}
\newcommand{\NestedDOneTwoIsolationTwoFSR}{-6.66}
\newcommand{\NestedDOneTwoCalibrationTargetPower}{11.8\%}
\newcommand{\NestedDOneTwoCalibrationOtherInBandPower}{70.3\%}
\newcommand{\NestedDOneTwoCalibrationOutOfBandPower}{17.8\%}
\newcommand{\NestedDOneTwoOOBMedianZero}{-10.1}
\newcommand{\NestedDOneTwoOOBMinZero}{-17.6}
\newcommand{\NestedDOneTwoOOBMaxZero}{-2.2}

\newcommand{\NestedDOneTwoMFourBlocking}{60.4\%}

}{}
\IfFileExists{results/joint_site_architecture_macros.tex}{

\newcommand{\ConverterPlaneZeroBlocking}{51.7\%}
\newcommand{\ConverterPlaneOneBlocking}{23.8\%}
\newcommand{\ConverterPlaneFullBlocking}{15.0\%}

\newcommand{\ConverterPlanePStar}{4}
\newcommand{\ConverterPlanePStarBranches}{32}
\newcommand{\ConverterPlanePStarTotalBlocks}{52}

}{}

\begin{document}

\title{Requirements for Synthetic-Frequency Switching in Integrated Photonic Fabrics}
\author{Jorge Parra}
\email{jorge.parra@uv.es}
\affiliation{Institute of Materials Science (ICMUV), University of Valencia, C/ Catedr\'atico Jos\'e Beltr\'an 2, 46980 Paterna, Valencia, Spain}
\date{November 20, 2026}

\begin{abstract}
Electro-optic synthetic-frequency coupling can coherently transfer light between the frequency modes of an integrated resonator, but its value in a multiport photonic switch depends on how connections share spatial paths, frequency channels, and control signals. Here we identify two distinct architectural roles of synthetic-frequency coupling in switching fabrics and the device requirements associated with each. As a channel-assignment resource, it cannot increase the number of orthogonal channels per port, but it recovers connections blocked by fixed input frequencies or unavailable output channels. A short coupling range is sufficient. In an $8\times8$ fabric with eight channels per port, coupling over the first three frequency spacings achieves \RevisionComplexityFRecAThreeQEight{} of the blocking reduction of unrestricted inter-mode coupling, and the range of coupled separations matters more than the specific choice of radio-frequency tones, consistent with limited-range wavelength conversion in wavelength-division-multiplexed networks. As a spatial-switching resource, frequency control can reduce the number of spatial switching elements only through a nonseparable joint element whose spatial state is programmable for each frequency channel, whereas a separate frequency-conversion stage cannot replace missing spatial connectivity. Finally, a thin-film lithium niobate resonator model with experimentally reported coupling and decay rates, embedded in a multistage Mach--Zehnder interferometer fabric, shows that with a single shared drive the power redistributed to non-target modes and the reduced transmission of frequency-preserving paths outweigh the added connectivity. These results define the through-channel transmission, mode selectivity, and control requirements for useful synthetic-frequency switching.
\end{abstract}


\maketitle
\section{Introduction}
\label{sec:introduction}

\begingroup
\sloppy
The rapid growth of data movement in datacenters, artificial-intelligence accelerators, and high-performance computing is increasing the demand for optical interconnects with high bandwidth density and low energy consumption \citep{rizzo2023massively,daudlin2025three,zhou2024silicon}. Optical circuit switches can complement their electrical counterparts by establishing high-capacity optical paths without repeated optical-to-electrical conversion \citep{patronas2025optical,farrington2010helios,mellette2017rotornet,mellette2020opera,liu2023lightwave,jouppi2023tpuv4}. More generally, programmable photonic circuits provide a route toward reconfigurable multiport optical functions using meshes of waveguides, tunable couplers, and phase shifters \citep{bogaerts2020programmable}. Scaling these systems requires routing not only across spatial ports but also across the frequency, polarization, and spatial-mode channels used to increase the bandwidth of each waveguide \citep{zhang2020multidimensional,liu2025roadm,wang2025roadm,liu2024selective}.
\par
\endgroup

When a switching architecture preserves the channel index along the optical path, a fixed source frequency, a channel-continuity requirement, or an unavailable output channel can block a connection even when another channel is free. The benefit of moving a connection to a different channel has been analyzed extensively for wavelength-division-multiplexed (WDM) networks and space--wavelength switches with wavelength converters \citep{sharony1992wsds,dasylva2002nonblocking,ngo2006wdm,hamza2007strict,chu2003placement,eramo2008shared,hamza2019conversioncomplexity}. In particular, limited-range conversion, in which an input wavelength can only be translated to nearby wavelengths, was shown to provide most of the blocking reduction of full-range conversion \citep{yates1996limited,ramaswami1998limited,tripathi2000limited}. These results are combinatorial and do not depend on the converter technology.

Synthetic dimensions provide a physical mechanism for this type of channel transfer within integrated photonics \citep{yuan2018synthetic,ehrhardt2023perspective,lustig2021topological}. In a dynamically modulated resonator, radio-frequency (RF) tones coherently couple modes separated by one or several multiples of the free spectral range (FSR), and the coupling strength, phase, and range are set by the modulation waveform \citep{yuan2021tutorial,dutt2019bands,dinh2024reconfigurable}. Integrated electro-optic synthetic-frequency lattices have been demonstrated on silicon complementary metal--oxide--semiconductor (CMOS) and thin-film lithium niobate (TFLN) platforms \citep{balcytis2022cmos,dinh2024reconfigurable}, together with long-range coupling, gauge potentials, and multidimensional synthetic spaces \citep{dutt2019bands,dutt2020twodim,cheng2023multidimensional}. Synthetic-frequency devices have also been proposed or demonstrated for programmable linear transformations, frequency-domain signal processing, and optical computing \citep{lukens2020frequencyprocessor,lu2020agile,buddhiraju2021arbitrary,zhao2022syntheticcomputing,wang2025programmable}. Here, \emph{synthetic-frequency coupling} refers to linear, coherent electro-optic coupling among resonator frequency modes, not to nonlinear frequency generation.

A synthetic-frequency converter differs from the idealized converters assumed in network analyses in three respects that matter for a multiport fabric. First, one resonator and one RF waveform define the coupling for every channel that traverses it, so several requested transfers must be compatible with a common control setting. Second, the transfer is coherent and multimode, so power can be redistributed to non-target channels inside and outside the switching band. Third, the frequency-preserving path through a driven resonator is itself modified by the drive. In addition, wavelength-selective spatial routing has been demonstrated with carrier-preserving resonant elements \citep{huang2020spacewavelength,zhang2025dilated}, and joint space--frequency transformations have been formulated for frequency-dependent programmable architectures \citep{friedman2025spacefrequency}, which raises the question of whether frequency control can also contribute to the spatial switching function itself.

In this work, we determine the architectural roles of synthetic-frequency coupling in an integrated photonic switching fabric and the device properties that each role requires. We restrict the analysis to switching, in which each connection occupies one output channel, and leave general multimode unitary processing in the synthetic frequency dimension \citep{buddhiraju2021arbitrary} outside the present scope. The work makes three contributions. (i) For frequency-channel switching, we quantify how the range and pattern of directly coupled mode separations determine connection blocking, and we relate the result to limited-range wavelength conversion. (ii) For spatial switching, we establish when frequency control can reduce the number of spatial switching elements and show that this requires a nonseparable joint space--frequency element rather than a separate conversion stage. (iii) For the physical device, we evaluate an electro-optically driven TFLN resonator embedded in a multistage Mach--Zehnder interferometer (MZI) network, using experimentally reported coupling and decay rates, and identify which device properties limit the fabric-level benefit. TFLN is used as a representative platform because it combines strong electro-optic interaction with low optical loss and high microwave bandwidth \citep{zhu2021tfln,hu2025integrated,dinh2024reconfigurable,dinh2026bounded}. The architectural analysis, by contrast, does not depend on the converter platform.

\section{Results}
\label{sec:results}

\subsection{Frequency-channel switching with synthetic-frequency coupling}
\label{sec:multiplexing-connectivity}
\label{sec:logical-equivalence-results}
\label{sec:state-constrained-scheduling}
\label{sec:conditional-scheduling-results}
\label{sec:connectivity-complexity}
\label{sec:connectivity-structure}
\label{sec:connectivity-programmability}

\begin{figure*}[t]
\centering
\includegraphics[width=\textwidth]{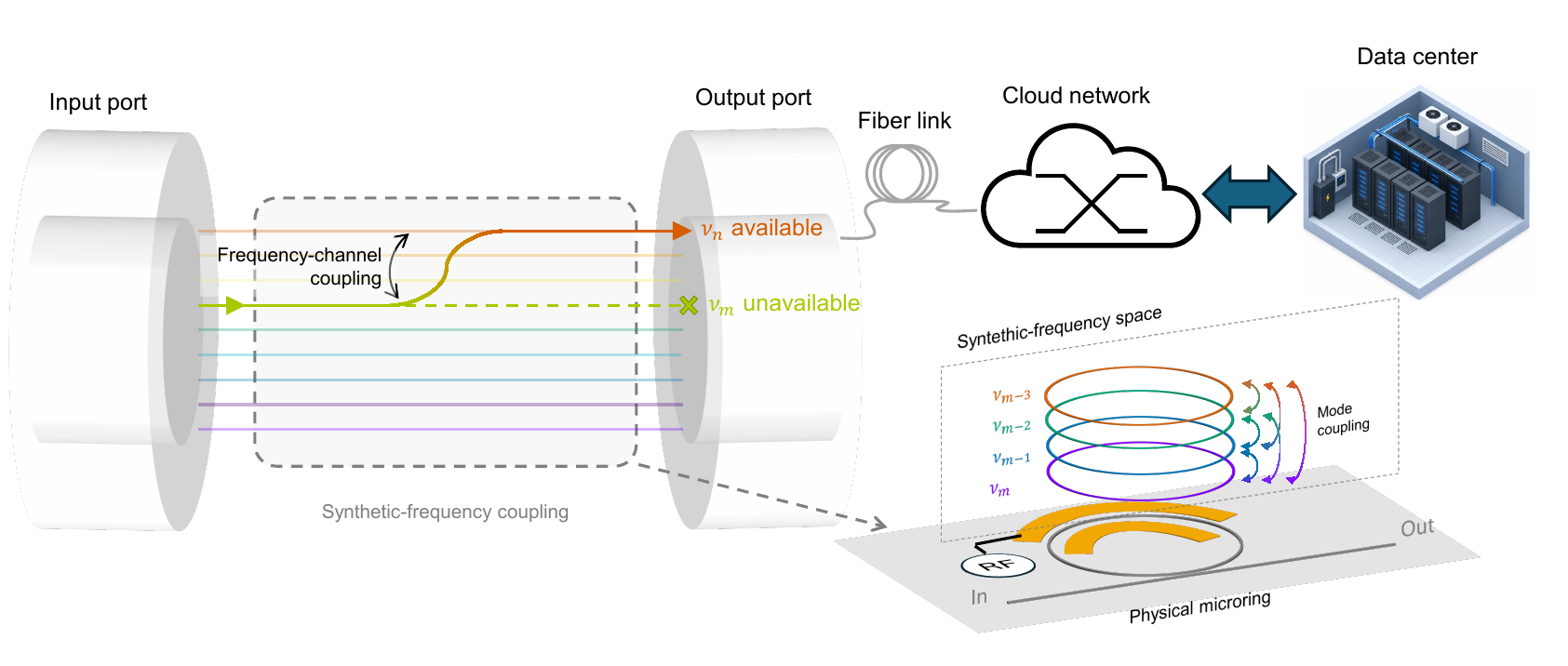}
\caption{
Synthetic-frequency coupling in a photonic switching fabric. Multiple frequency channels share each physical input and output port. When the output channel corresponding to input mode $\nu_m$ is unavailable, synthetic-frequency coupling transfers the signal to an available mode $\nu_n$ before transmission over the external optical link. The inset maps the RF-driven physical microring resonator onto coupled modes in the synthetic-frequency dimension.
}
\label{fig:architectures}
\end{figure*}

\Cref{fig:architectures} summarizes the switching function considered here. A fabric has $N$ physical input and output ports, each carrying $N_{\mathrm{ch}}$ orthogonal frequency channels. If the output channel corresponding to a fixed input frequency is unavailable, coherent frequency-mode coupling can transfer the signal to an available output channel. Optical multiplexing provides the channels in parallel, and synthetic-frequency coupling adds controlled off-diagonal coupling between them.

A counting argument first bounds where a benefit can arise. Let $N_{\mathrm{conn}}(p)$ be the number of connections associated with physical port $p$, and $N_{\mathrm{conn}}^{\max}=\max_p N_{\mathrm{conn}}(p)$. If all channels can be freely selected and independently routed, the minimum number of switching configurations required to establish all connections is the same with independent frequency-channel switching ($N_{\mathrm{config}}^{\mathrm{mux}}$) and with ideal synthetic-frequency coupling ($N_{\mathrm{config}}^{\mathrm{syn}}$),
\begin{equation}
N_{\mathrm{config}}^{\mathrm{syn}}
=N_{\mathrm{config}}^{\mathrm{mux}}
=\left\lceil\frac{N_{\mathrm{conn}}^{\max}}{N_{\mathrm{ch}}}\right\rceil
\label{eq:flexible-equivalence-results}
\end{equation}
(Supplementary Note~1). Inter-mode coupling changes which input--output channel mappings are possible, but it does not add channels to a port. Its benefit can only appear when the channel of a connection is constrained, which is the situation analyzed in the remainder of this section.

We consider an idealized fabric with $N=8$ ports, $N_{\mathrm{ch}}=8$ channels, and 32 simultaneous connections. The input port, output port, and input frequency of each connection are fixed, input port--frequency pairs are sampled without replacement, and a fraction of the output channels is unavailable. Routes and output channels are assigned jointly by an exact mixed-integer program that maximizes the number of established connections. Frequency-preserving routing, restricted coupling, and unrestricted inter-mode coupling are compared on the same switching instances (30 seeds). Because this model contains no resonator loss, crosstalk, or conversion penalty, it isolates the combinatorial effect of the allowed transfers. The protocol is summarized in \cref{sec:methods-optical-switching-architecture} and Supplementary Table~S1.

\begin{figure}[!tbp]
\centering
\includegraphics[width=0.92\columnwidth]{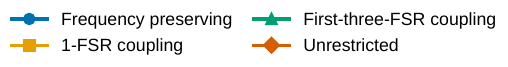}\\[-3.5ex]
\subfloat[]{%
\includegraphics[width=0.45\columnwidth]{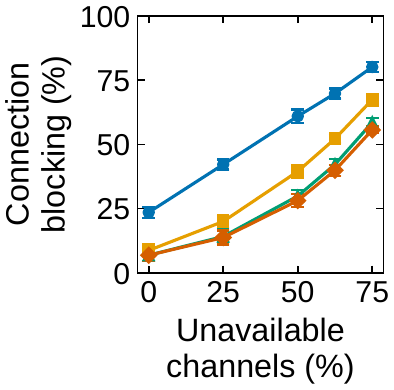}}\hfill
\subfloat[]{%
\includegraphics[width=0.45\columnwidth]{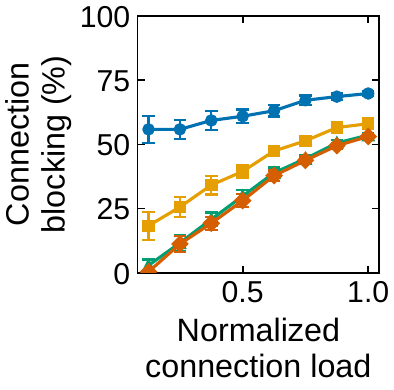}}
\caption{
Idealized frequency-channel switching. Connection blocking versus (a) output-channel unavailability and (b) normalized connection load at 50\% unavailability, for frequency-preserving routing, restricted synthetic-frequency coupling, and unrestricted inter-mode coupling. Points show means over 30 seeds with 95\% confidence intervals.
}
\label{fig:global-pinned}
\end{figure}

At 50\% output-channel unavailability, frequency-preserving routing blocks \GlobalBaseline{} of the connections. Nearest-neighbor coupling reduces blocking to \GlobalRFOne{}, coupling over the first three frequency spacings to \GlobalRFThree{}, and unrestricted coupling to \GlobalFull{} [\cref{fig:global-pinned}(a)]. The same ordering holds across connection loads [\cref{fig:global-pinned}(b)].

To compare different coupling sets, we normalize the blocking reduction of each set to that of unrestricted coupling on the same instance, and we define the utilization of separation $\ell$ as
\begin{equation}
U_\ell=\frac{N_{\rm est}^{(\ell)}}{N_{\rm est}},
\qquad
N_{\rm est}^{(\ell)}=N_{\rm est}(|n-m|=\ell),
\label{eq:synthetic-frequency-utilization}
\end{equation}
where $m$ and $n$ are the input and output mode indices and $N_{\rm est}^{(\ell)}$ is the number of established connections that use separation $\ell$. $U_0$ is the frequency-preserving share.

\begin{figure*}[t]
\centering
\subfloat[]{%
\includegraphics[width=0.31\textwidth]{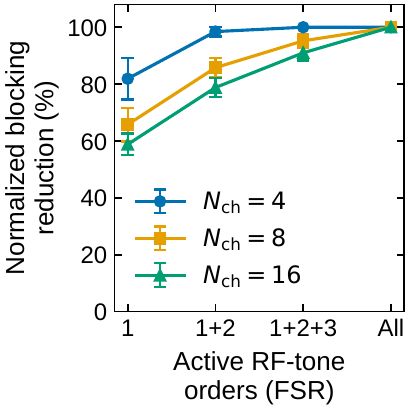}}\hfill
\subfloat[]{%
\includegraphics[width=0.31\textwidth]{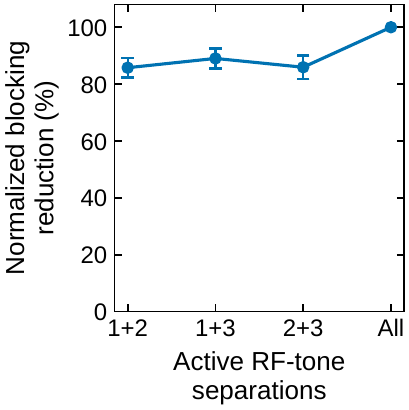}}\hfill
\subfloat[]{%
\includegraphics[width=0.31\textwidth]{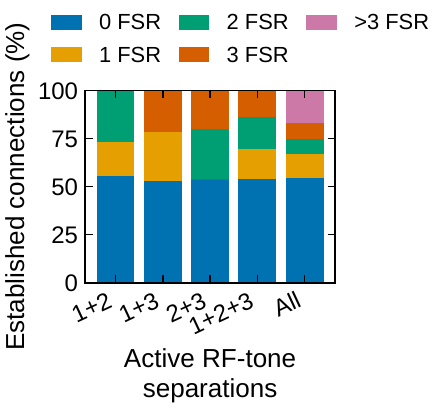}}
\caption{
Coupling range and coupling pattern. (a) Blocking reduction normalized to unrestricted coupling for $N_{\mathrm{ch}}=4$, 8, and 16 as RF tones at successive FSR orders are added. (b) Normalized blocking reduction for different two-tone configurations at $N_{\mathrm{ch}}=8$. (c) Fraction of established connections using each mode separation. All panels use a normalized load of 0.5 and 50\% output-channel unavailability, and error bars indicate 95\% confidence intervals over 30 seeds.
}
\label{fig:connectivity-complexity}
\end{figure*}

For $N_{\mathrm{ch}}=8$, RF tones at the first three FSR orders recover \RevisionComplexityFRecAThreeQEight{} (95\% confidence interval $\pm$\RevisionComplexityFRecAThreeQEightCI{} percentage points) of the blocking reduction obtained with unrestricted coupling [\cref{fig:connectivity-complexity}(a)]. The corresponding values are \RevisionComplexityFRecAThreeQFour{} for $N_{\mathrm{ch}}=4$ and \RevisionComplexityFRecAThreeQSixteen{} for $N_{\mathrm{ch}}=16$, so a fixed range covers a decreasing fraction of the benefit as the number of channels grows, and the range of directly coupled separations sets most of the result. Averaged over the broader set of loads and channel-fragmentation levels in Supplementary Note~4, the three-order value is \RevisionFRecAThreeQEight{} for $N_{\mathrm{ch}}=8$ and \RevisionFRecAThreeQSixteen{} for $N_{\mathrm{ch}}=16$. This behavior agrees with WDM network analyses, in which limited-range conversion captures most of the benefit of full-range conversion \citep{yates1996limited,ramaswami1998limited,tripathi2000limited}. The same conclusion thus holds in the synthetic-frequency setting, where the RF tone orders set the transfer range.

By comparison, the choice of separations at a fixed number of RF tones has a minor effect. For $N_{\mathrm{ch}}=8$, tones at one and two, one and three, and two and three FSRs give \RevisionFixedCostBOneTwoBlocking{}, \RevisionFixedCostBOneThreeBlocking{}, and \RevisionFixedCostBTwoThreeBlocking{} blocking, respectively [\cref{fig:connectivity-complexity}(b)], and the paired differences remain below one percentage point and within their 95\% confidence intervals. Nonconsecutive separations are nevertheless used when available. With tones at one and three FSRs, for example, \RevisionFixedCostBOneThreeUtilizationOne{} and \RevisionFixedCostBOneThreeUtilizationThree{} of the established connections use one- and three-FSR transfers, respectively [\cref{fig:connectivity-complexity}(c)]. Reconfiguring the tone pair for each switching instance provides only a modest additional gain (Supplementary Note~4).

\subsection{Joint space--frequency switching and spatial switching hardware}
\label{sec:spatial-factorization-boundary}
\label{sec:nonseparable-spatial-substitution}

Frequency control could also reduce the number of spatial switching elements. To examine this possibility, let $\Pi_1$ and $\Pi_2$ be frequency-independent spatial switching stages, $U_\nu$ a frequency-only coupling transformation, and $I_N$ and $I_\nu$ the identities in the spatial and frequency-channel spaces. A cascade of these elements remains separable,
\begin{equation}
(\Pi_2\otimes I_\nu)(I_N\otimes U_\nu)(\Pi_1\otimes I_\nu)
=(\Pi_2\Pi_1)\otimes U_\nu ,
\label{eq:space-state-factorization}
\end{equation}
so the set of reachable output ports does not depend on the input frequency. A separate frequency-conversion stage can give access to an otherwise unavailable channel, but it cannot create a spatial path that the spatial network does not provide (Supplementary Note~1).

Frequency-dependent spatial routing instead requires each $2\times2$ switching element, indexed by its Benes stage $s=1,\ldots,2\log_2(N)-1$ and row $r=1,\ldots,N/2$ for $N=2^k$, to implement the nonseparable response
\begin{equation}
\mathbf{S}_{\mathrm{sf}}^{(s,r)}=\sum_{m,n=1}^{N_{\mathrm{ch}}}
\mathbf{S}_{nm}^{(s,r)}\otimes|\nu_n\rangle\langle\nu_m|,
\qquad \mathbf{S}_{nm}^{(s,r)}\in\mathbb C^{2\times2},
\label{eq:joint-space-frequency-site}
\end{equation}
where $|\nu_n\rangle\langle\nu_m|$ maps input mode $\nu_m$ to output mode $\nu_n$ and $\mathbf{S}_{nm}^{(s,r)}$ is the corresponding $2\times2$ spatial response. With $a,b\in\{1,2\}$ labeling the local ports, the amplitude from $|a,\nu_m\rangle$ to $|b,\nu_n\rangle$ is $[\mathbf{S}_{nm}^{(s,r)}]_{ba}$. Independently programmable diagonal blocks $\mathbf{S}_{mm}^{(s,r)}$ let different channels select different bar or cross states while keeping their carrier, and off-diagonal blocks ($n\ne m$) add frequency conversion along the selected route.

The consequence for hardware can be quantified for a fabric in which each of the $N_{\mathrm{ch}}$ channels supports an independent input--output port mapping and every port carries the full channel grid. A binary $N\times N$ Benes network contains
\begin{equation}
N_{\rm sw}(N)=\frac{N}{2}\left(2\log_2(N)-1\right)
\label{eq:benes-spatial-sites}
\end{equation}
$2\times2$ elements. Counting elements without regard to their internal complexity, parallel per-channel networks and an ideal joint network require
\begin{equation}
N_{\rm sw}^{\rm ind}=N_{\mathrm{ch}}N_{\rm sw}(N),
\qquad
N_{\rm sw}^{\rm joint}=N_{\rm sw}(N),
\label{eq:spatial-site-counts}
\end{equation}
that is, a reduction
\begin{equation}
R_{\rm sw}=1-\frac{N_{\rm sw}^{\rm joint}}{N_{\rm sw}^{\rm ind}}
=1-\frac{1}{N_{\mathrm{ch}}}.
\label{eq:spatial-site-reduction}
\end{equation}
This element count, used throughout this section, does not include insertion loss, footprint, or control channels, which are discussed in \cref{sec:discussion}. The reduction in \cref{eq:spatial-site-reduction} relies only on the independently programmable diagonal blocks of \cref{eq:joint-space-frequency-site}, not on frequency conversion. A frequency-independent Benes network followed by a separate conversion stage has the same count $N_{\rm sw}(N)$ but, by \cref{eq:space-state-factorization}, cannot route channels independently and is consequently not functionally equivalent to the parallel networks.

\begin{figure*}[tbp]
\centering
\subfloat[]{%
\includegraphics[width=0.32\textwidth]{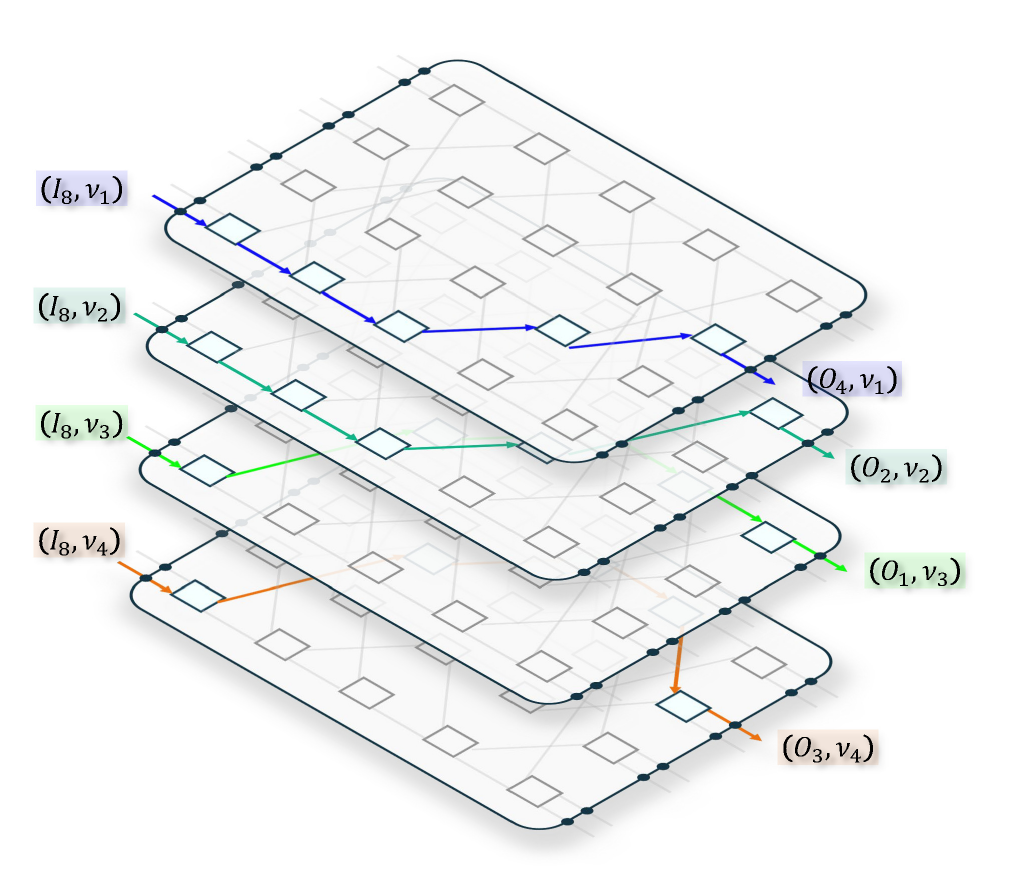}}\hfill
\subfloat[]{%
\includegraphics[width=0.32\textwidth]{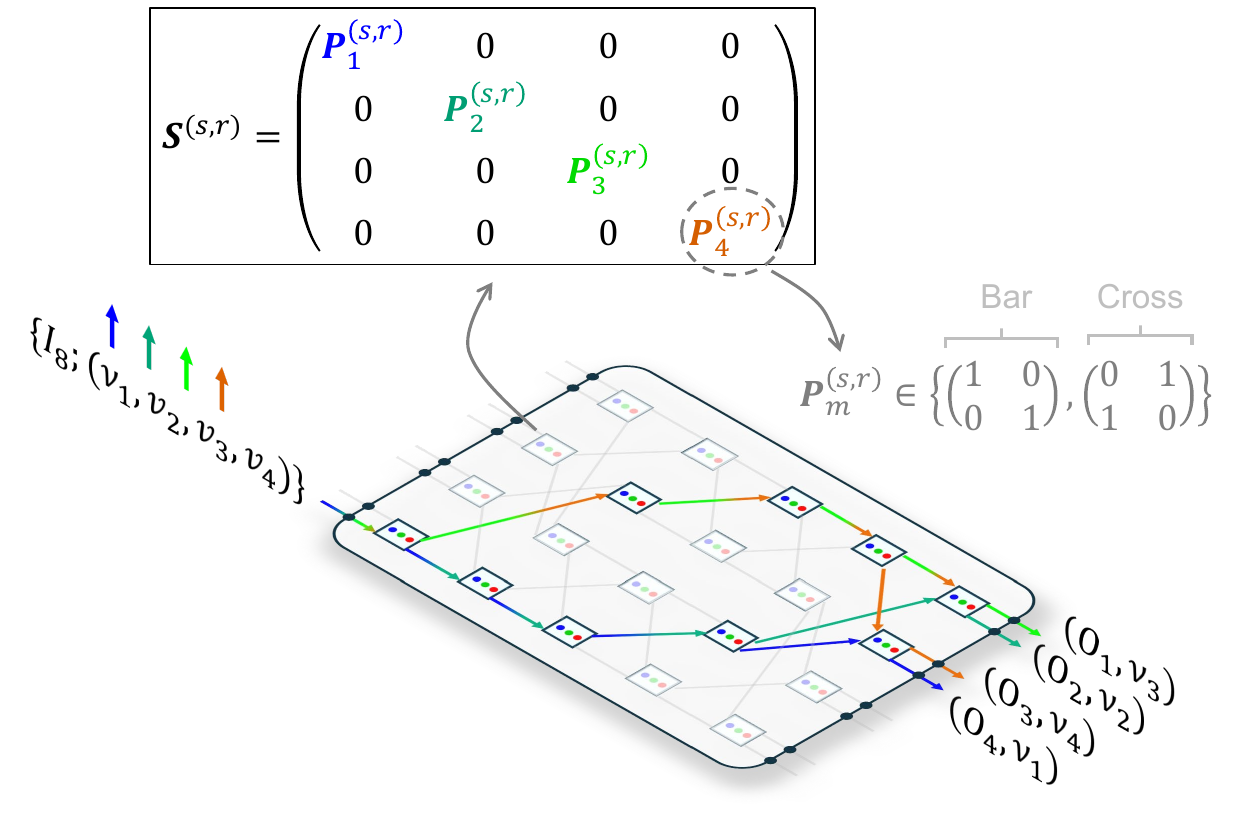}}\hfill
\subfloat[]{%
\includegraphics[width=0.32\textwidth]{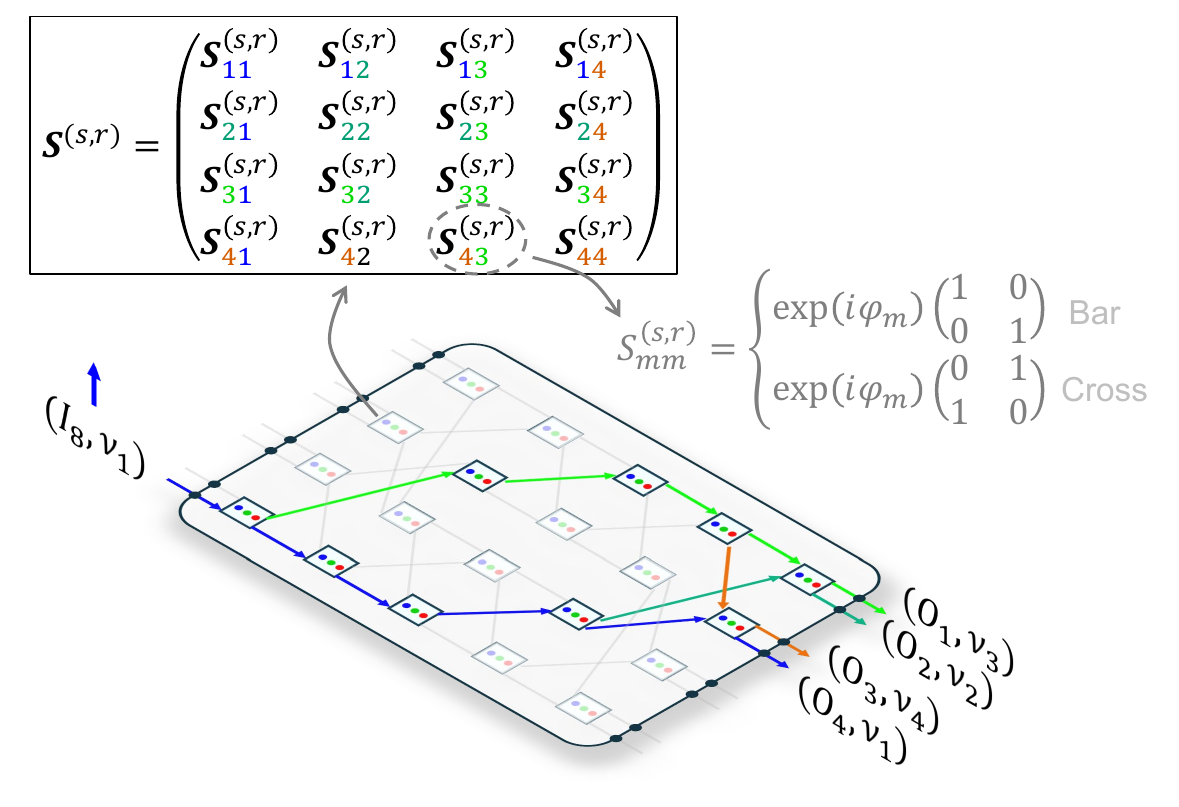}}
\caption{
Parallel and joint space--frequency Benes architectures for $N=8$ and $N_{\mathrm{ch}}=4$. (a) Independent frequency-preserving Benes networks, one per channel. (b) Joint architecture with frequency-dependent bar/cross states and preserved carriers. (c) Joint architecture including off-diagonal blocks for frequency conversion.
}
\label{fig:space-state-boundary}
\end{figure*}

\begin{figure}[tbp]
\centering
\subfloat[]{%
\includegraphics[width=0.48\columnwidth]{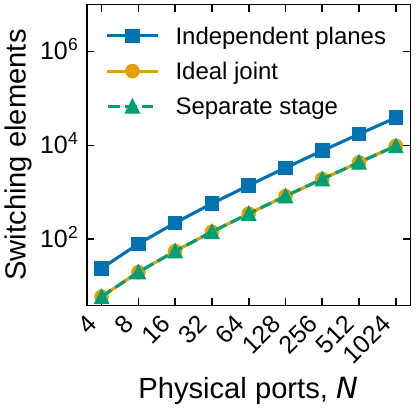}}\hfill
\subfloat[]{%
\includegraphics[width=0.48\columnwidth]{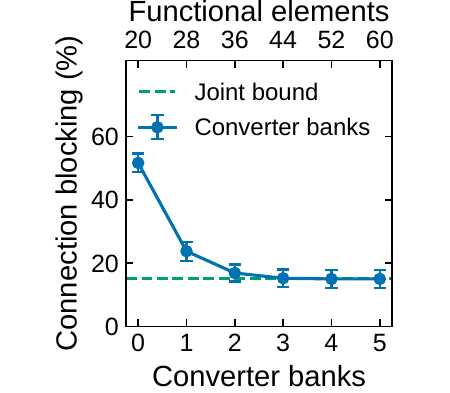}}
\caption{
Spatial element count and converter placement. (a) Number of spatial switching elements versus port count for independent per-channel networks, the ideal joint architecture, and a separate frequency-conversion stage ($N_{\mathrm{ch}}=4$). (b) Connection blocking in an $8\times8$ five-stage frequency-selective Benes network as independently controlled converter banks are added. The dashed line indicates the ideal joint space--frequency result, and error bars show 95\% confidence intervals.
}
\label{fig:spatial-element-scaling}
\end{figure}

\Cref{fig:space-state-boundary} illustrates these relations for $N=8$ and $N_{\mathrm{ch}}=4$. Four parallel Benes networks provide independent mappings for the four channels [\cref{fig:space-state-boundary}(a)], while the joint architecture shares one topology with independently programmable diagonal blocks [\cref{fig:space-state-boundary}(b)] and, optionally, off-diagonal blocks for conversion selected by the spatial path [\cref{fig:space-state-boundary}(c)]. An $8\times8$ Benes network has $N_{\rm sw}=20$ elements, so the parallel networks require 80 and the joint architecture 20 joint elements, a 75\% reduction in element count [\cref{fig:spatial-element-scaling}(a)].

Off-diagonal blocks become useful when channel availability varies along a lightpath. A cascaded alternative places banks of independently controlled per-waveguide frequency converters after selected Benes stages. Let $L_{\rm B}=2\log_2(N)-1$ be the number of stages and $\mathcal F_P$ the set of feasible assignments with $P$ nested converter banks. Because every converter can be bypassed, adding a bank never removes a feasible assignment. If each lightpath may undergo at most one ideal conversion after a selected element, a bank after every stage reproduces any allowed joint conversion event, and conversely each separate conversion maps onto an off-diagonal block. Hence
\begin{equation}
\begin{aligned}
\mathcal F_0&\subseteq\mathcal F_1\subseteq\cdots\subseteq
\mathcal F_{L_{\rm B}}=\mathcal F_{\rm joint}^{(1)},\\
B_{\rm joint}^{(1)}&=B_{L_{\rm B}}\leq\cdots\leq B_1\leq B_0,
\end{aligned}
\label{eq:converter-plane-inclusion}
\end{equation}
where $B$ denotes connection blocking (constructive proof in Supplementary Note~1). The equality holds for this single-conversion model, not for an arbitrary coherent matrix of the form of \cref{eq:joint-space-frequency-site}.

We evaluate this hierarchy for a frequency-selective $8\times8$ five-stage Benes network with $N_{\mathrm{ch}}=4$, 16 connections, and 25\% internal channel unavailability. Without conversion, blocking is \ConverterPlaneZeroBlocking{} [\cref{fig:spatial-element-scaling}(b)]. The best first bank, selected on calibration instances, reduces blocking on the evaluation instances to \ConverterPlaneOneBlocking{}, and further banks reduce it monotonically. With \ConverterPlanePStar{} banks, the cascaded architecture reproduces the joint result of \ConverterPlaneFullBlocking{} for every evaluation seed, at the cost of 20 frequency-selective elements plus \ConverterPlanePStarBranches{} independently controlled converters (\ConverterPlanePStarTotalBlocks{} functional elements), compared with 20 joint elements.

\subsection{Synthetic-frequency switching on a thin-film lithium niobate platform}
\label{sec:tfln-realization}
\label{sec:tfln-platform}
\label{sec:coherent-switching-results}

\begin{figure*}[!t]
\centering
\subfloat[]{%
\includegraphics[width=0.92\textwidth]{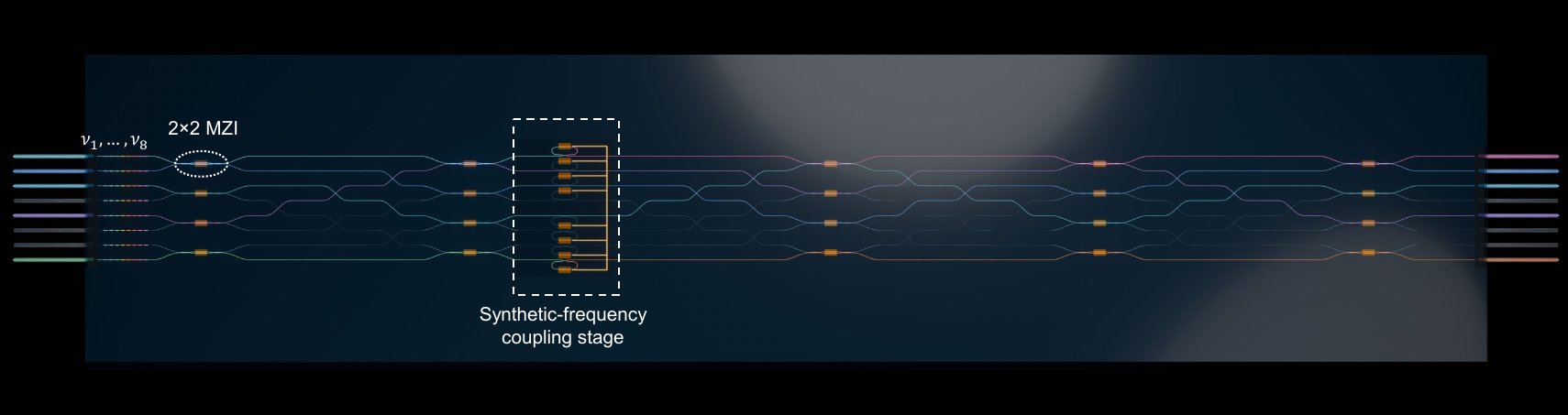}}

\subfloat[]{%
\includegraphics[height=0.177\textwidth]{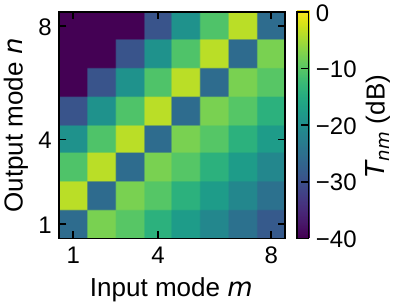}}\hfill
\subfloat[]{%
\includegraphics[height=0.25\textwidth]{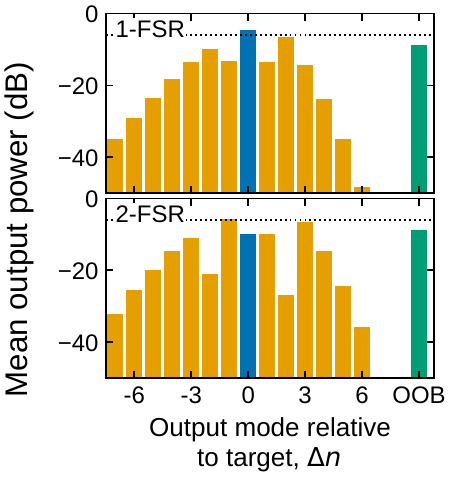}}\hfill
\subfloat[]{%
\includegraphics[height=0.25\textwidth]{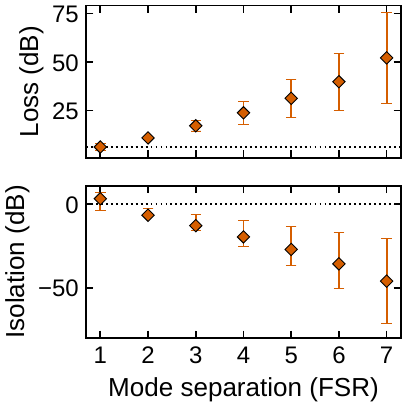}}

\subfloat[]{%
\includegraphics[height=0.25\textwidth]{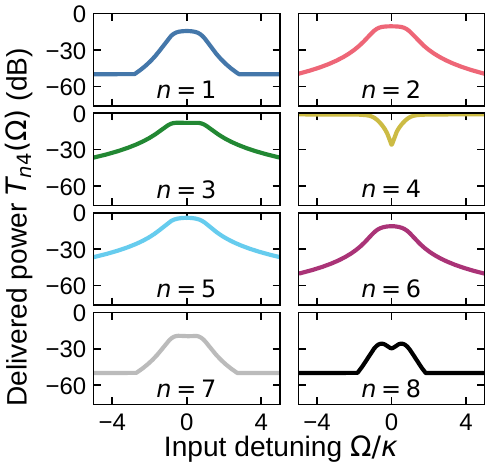}}\hfill
\subfloat[]{%
\includegraphics[height=0.25\textwidth]{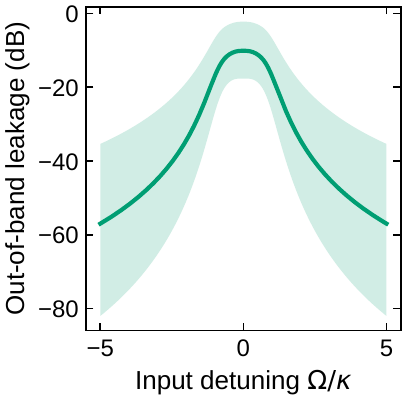}}\hfill
\subfloat[]{%
\includegraphics[height=0.25\textwidth]{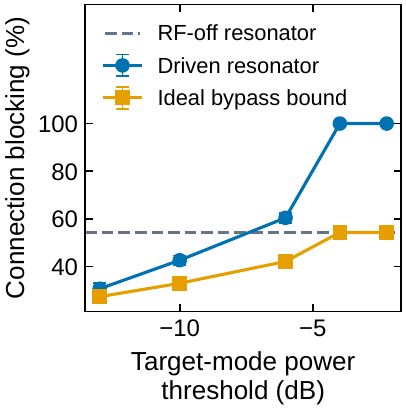}}

\caption{
Mode-resolved response and fabric-level performance of the modeled TFLN synthetic-frequency stage. (a) Multistage spatial network with an interstage resonator plane. (b) Mode-to-mode power-transfer matrix $T_{nm}$ for the optimized 1+2-FSR drive. (c) Output power in the target mode (blue), the other in-band modes (orange), and all out-of-band modes (green) for the calibration transfers used to optimize the 1+2-FSR drive. The dotted line marks the $-6$~dB threshold. (d) Conversion loss and in-band mode isolation versus mode separation. Diamonds show medians, error bars span the minimum-to-maximum range, and dotted lines mark 6~dB loss and 0~dB isolation. (e) Delivered power versus optical detuning for input mode $m=4$. (f) Out-of-band leakage versus detuning. The line is the median over the eight input modes and the shaded region spans the minimum-to-maximum range. (g) Connection blocking versus target-power threshold for the driven resonator, the RF-off resonator, and the ideal bypass bound. Markers show means over 30 seeds with 95\% confidence intervals, and the dashed line shows the mean RF-off reference.
}
\label{fig:tfln-optimized-topology}
\label{fig:photonic-switching-fabric}
\end{figure*}

We now replace the ideal transfers by the response of an electro-optically driven TFLN resonator placed between the second and third stages of a frequency-independent spatial MZI network [\cref{fig:photonic-switching-fabric}(a)]. The resonator is described by temporal coupled-mode theory (TCMT), including photon decay, external coupling, and interference between coupling pathways, with coupling and decay rates taken from an experimentally reported device \citep{dinh2024reconfigurable}. RF tones at one, two, and three times the FSR couple modes separated by the corresponding number of spacings \citep{yuan2021tutorial,dinh2024reconfigurable}. For each set of active tones, the coupling strengths and phases are optimized on calibration instances to increase the power delivered to the requested target modes and to suppress non-target modes. The resulting configuration is then held fixed and shared by all resonators, which form a single RF-control group. Model, optimization, and routing details are given in \cref{sec:methods-tfln-converter,sec:methods-routing-evaluation} and Supplementary Notes~2 and~3.

For input mode $m$, $T_{nm}$ is the fraction of optical power delivered to output mode $n$ [\cref{eq:methods-target-power}]. A transfer, including the frequency-preserving case $n=m$, is accepted when $T_{nm,\mathrm{dB}}=10\log_{10}T_{nm}\geq T_{\min,\mathrm{dB}}$. The reference threshold $T_{\min,\mathrm{dB}}=-6$~dB corresponds to 25\% delivered power after a 1-dB resonator-plane path loss. It acts as a screening criterion rather than a receiver specification, and the full threshold dependence is shown below. The reference calculation assumes $\eta_{\mathrm{ext}}=1$ (no intrinsic resonator loss), which is optimistic relative to the reported device. An ideal lossless bypass of the resonator for frequency-preserving paths is included as an upper bound.

\label{sec:tfln-fixed-control}
\label{sec:tfln-optimized-topology}

With RF tones at one and two times the FSR (the 1+2-FSR drive), the optimized response is dominated by $J_1$. Coherent coupling among the resonator modes then produces a multimode response rather than two independent frequency translations, and converted power also appears at separations that are not directly driven [\cref{fig:tfln-optimized-topology}(b)]. Averaged over the calibration transfers, the target mode receives \NestedDOneTwoCalibrationTargetPower{} of the output power, other in-band modes \NestedDOneTwoCalibrationOtherInBandPower{}, and out-of-band modes \NestedDOneTwoCalibrationOutOfBandPower{} [\cref{fig:tfln-optimized-topology}(c)]. One-FSR transfers show a median conversion loss of \NestedDOneTwoLossOneFSR{}~dB and a median target-to-strongest-non-target in-band isolation (mode isolation) of \NestedDOneTwoIsolationOneFSR{}~dB, whereas two-FSR transfers show \NestedDOneTwoLossTwoFSR{}~dB and $\NestedDOneTwoIsolationTwoFSR$~dB [\cref{fig:tfln-optimized-topology}(d)], consistent with the smaller optimized $J_2$ and with indirect pathways that redistribute power among non-target modes.

In addition, the response depends strongly on the input mode. At zero detuning, the median out-of-band leakage over the eight input modes is $\NestedDOneTwoOOBMedianZero$~dB, ranging from $\NestedDOneTwoOOBMinZero$ to $\NestedDOneTwoOOBMaxZero$~dB [\cref{fig:tfln-optimized-topology}(e,f)]. For switching, the conversion efficiency of a selected mode pair is not a sufficient description, because the full mode-resolved response determines how much power reaches the target channel and how much is redistributed.

At the fabric level [\cref{fig:tfln-optimized-topology}(g)], the driven resonator gives \NestedDOneTwoBlocking{} blocking at $T_{\min,\mathrm{dB}}=-6$~dB, compared with \NestedRFOffBlocking{} for the RF-off resonator and \NestedBypassBoundBlocking{} for the ideal bypass bound. With the drive on, the diagonal terms $T_{mm}$ fall below the acceptance threshold, so connections that did not need conversion are lost, and the added conversion paths do not compensate for this loss. The gap between the driven and bypass curves quantifies the benefit that is lost to this effect. Adding a third tone at three times the FSR does not improve the result within the explored control range, because the optimizer suppresses it ($J_3=\NestedRichJThree$~GHz). These RF settings are bounded local-search solutions rather than global optima. Raising the architectural assignment limit from $n_{\mathrm{sim}}=4$ to 8 leaves blocking unchanged at \NestedDOneTwoMFourBlocking{}, so this limit is not active (Supplementary Note~4).

\section{Discussion}
\label{sec:discussion}

\subsection{Two architectural roles}

The results separate two roles of synthetic-frequency coupling in a switching fabric. In the first, frequency conversion relaxes channel constraints by moving a connection from its input carrier to another available channel. In the second, a frequency-dependent spatial response lets different channels follow independent routes within one switching topology. The first role improves access to existing channels, whereas the second changes the spatial switching function.

For the first role, \cref{eq:flexible-equivalence-results} shows that coupling cannot add orthogonal channels to a port, so any benefit comes from relaxing constraints on channel assignment. Within that regime, the range of directly coupled separations largely determines the achievable blocking reduction, and different two-tone patterns perform within statistical uncertainty of each other (\cref{fig:connectivity-complexity}). This is consistent with network analyses of limited-range wavelength conversion \citep{yates1996limited,ramaswami1998limited,tripathi2000limited}. The required range must be matched to the number of channels and to their availability constraints, and in a resonant implementation it is set by the highest RF tone order that can be applied with sufficient coupling strength.

For the second role, \cref{eq:space-state-factorization} shows that a conversion stage between frequency-independent spatial stages remains separable, so it can change the carrier but cannot create a spatial path. A reduction in spatial element count consequently requires the nonseparable response of \cref{eq:joint-space-frequency-site}, and the reduction in \cref{eq:spatial-site-reduction} comes from its independently programmable diagonal blocks. The converter-bank analysis [\cref{fig:spatial-element-scaling}(b)] shows the corresponding trade-off. Separate converters can reproduce the single-conversion joint function, but only if converter banks are distributed through the network. Because this is a comparison of functional element counts, a practical joint element must be assessed by its insertion loss, footprint, electrical and RF control, and calibration effort. Frequency-dependent spatial switching has been demonstrated with carrier-preserving microring-assisted MZI and resonant crosspoint fabrics \citep{huang2020spacewavelength,zhang2025dilated}. Synthetic-frequency coupling would add controlled transfer between channels to these elements, and a practical design should be compared with both independent per-channel planes and frequency-selective fabrics followed by separate converters.

\subsection{Device requirements}

The TFLN calculation translates the architectural benefits into device requirements, namely sufficient power in the target mode, high transmission of frequency-preserving paths, and adequate mode selectivity. Efficient transfer in a dynamically modulated resonator requires the electro-optic coupling rate to be comparable to the photon-decay rate \citep{fan2003tcmt,dinh2024reconfigurable}. With the reported scales and $\kappa=2\gamma=1.26$~GHz, $J_1/\kappa=0.27$, $J_2/\kappa=0.19$, and $J_3/\kappa=0.17$, which places the reference device in the regime of significant coherent redistribution among modes. TFLN is well suited to increasing these ratios because of its low optical loss, strong electro-optic response, and high microwave bandwidth \citep{zhu2021tfln,hu2025integrated}.

Stronger coupling alone is not sufficient, however. In the modeled device, most of the output power remains in non-target in-band modes or leaves the switching band, and the drive degrades the frequency-preserving paths enough to increase blocking relative to the undriven resonator. The comparison with the bypass bound in \cref{fig:tfln-optimized-topology}(g) indicates that the largest single improvement would come from not routing frequency-preserving connections through a driven resonator, for example through a switchable bypass or by detuning unused resonators. Beyond that, the requirements are higher mode selectivity, so that a transfer above threshold is not accompanied by comparable power in other channels, and enough independent control to accommodate simultaneous transfers. Additional RF-control groups improve routing flexibility but increase RF routing, phase control, calibration effort, and sensitivity to resonator detuning and thermal drift. Resonant conversion is also only one option, since single-pass and dual-parallel MZI modulators and cascaded electro-optic modulators with spectral shaping offer different trade-offs in efficiency, bandwidth, loss, footprint, and control \citep{kodigala2019ssb,lukens2020frequencyprocessor,lu2020agile}.

Experimental validation should target the complete complex multimode scattering matrix $S_{nm}$ under the RF amplitudes and phases used for switching. Replacing the TCMT response by measured matrices would allow the same routing framework to evaluate blocking, delivered power, conversion loss, mode isolation, and spurious sidebands for a real device.

\subsection{Scope and limitations}
\label{sec:limitations}

Several modeling choices limit the scope of these conclusions.

\emph{Routing model.} Blocking is computed for static sets of simultaneous connection requests with exact assignment, without dynamic arrivals, rearrangement costs, or contention between successive configurations. Connections are admitted on the delivered power in the target mode. Power scattered into channels used by other connections is reported as mode isolation and out-of-band leakage, but it does not enter the admission criterion. The median two-FSR mode isolation of $\NestedDOneTwoIsolationTwoFSR$~dB means that the strongest non-target channel typically receives more power than the target channel. Any crosstalk requirement added to the admission criterion could therefore only remove transfers, so the fabric-level TFLN blocking values are lower bounds on the blocking obtained when crosstalk is enforced.

\emph{Control model.} In the TFLN calculation, the RF amplitudes and phases are calibrated once per tone set on calibration instances and then shared by all resonators and held fixed for all evaluation instances. In effect, the resonator plane acts as a fixed coherent scattering element, and switching is performed by the spatial network and the channel assignment. Per-request RF reconfiguration, several RF-control groups, and a switchable bypass are not part of the reference model, although the ideal bypass bound indicates the margin available from the last of these.

\emph{Signal bandwidth and channel grid.} In a resonant implementation the channel spacing equals the resonator FSR, and the signal bandwidth that can traverse the resonator without filtering is limited by the resonance linewidth, here $\kappa=1.26$~GHz. The reference parameters thus describe narrowband channels on a grid set by the resonator, not the 50--200~GHz dense-WDM grid with high-symbol-rate signals. Matching such grids would require resonators with correspondingly larger FSR and linewidth, together with proportionally larger coupling rates to maintain $J_\ell/\kappa$, or nonresonant conversion schemes.

\emph{Idealized architectural results.} The element counts of \cref{eq:spatial-site-counts,eq:spatial-site-reduction} assume ideal joint elements and do not account for loss or control complexity. The idealized blocking results use a specific traffic model (Supplementary Table~S1). Although the qualitative conclusions are stable across the loads and fragmentation levels tested, the exact fractions depend on the workload.

\section{Conclusions}
\label{sec:conclusions}

We have identified two architectural roles of synthetic-frequency coupling in integrated photonic switching fabrics and the device properties that each requires. As a channel-assignment resource, synthetic-frequency coupling does not add channels to a port but recovers connections blocked by frequency constraints. A short coupling range is sufficient. In an $8\times8$ fabric with eight channels per port, coupling over the first three frequency spacings captures \RevisionComplexityFRecAThreeQEight{} of the blocking reduction of unrestricted coupling, and the coupling range matters more than the specific choice of RF tones, in agreement with limited-range wavelength conversion in WDM networks. As a spatial-switching resource, frequency control can reduce the number of spatial switching elements only through a nonseparable joint element whose spatial state is programmable per channel, while a separate conversion stage cannot replace missing spatial connectivity.

At the device level, a TFLN resonator model with experimentally reported coupling and decay rates shows that additional mode connectivity is useful only when accompanied by high through-channel transmission and mode selectivity. With a single shared and fixed drive, the redistribution of power to non-target modes and the degradation of frequency-preserving paths outweigh the added connectivity. Bypassing the resonator for unconverted connections, higher mode isolation, and independent control are the main device targets. Future work should measure complete complex scattering matrices under switching conditions, include crosstalk in the admission criterion, and evaluate per-request RF control. More broadly, the framework applies to programmable photonic systems that combine spatial and spectral degrees of freedom, including frequency-domain signal processing, linear transformations, and photonic computing.

\section{Methods}
\label{sec:methods}

\subsection{Switching model and frequency-channel constraints}
\label{sec:methods-optical-switching-architecture}

We consider a photonic switching fabric with $N$ input and output ports and $N_{\mathrm{ch}}$ frequency channels per port. Each optical connection is defined by an input port, an output port, and a fixed input frequency. Spatial routes and output frequency channels are selected jointly to maximize the number of established connections, subject to channel availability and single-channel occupancy.

The allowed frequency transfer is determined by the set of active RF-tone orders, $\mathcal{I}_{\mathrm{RF}}$. An input mode $m$ can be connected to an output mode $n$ when
\begin{equation}
m=n
\qquad \text{or} \qquad
|m-n|\in\mathcal{I}_{\mathrm{RF}} .
\label{eq:ideal-frequency-connectivity}
\end{equation}
Here, $\mathcal{I}_{\mathrm{RF}}=\varnothing$ corresponds to frequency-preserving routing, while finite sets describe restricted synthetic-frequency coupling. Unrestricted inter-mode coupling is used as an ideal reference.

The main idealized study uses $N=8$, $N_{\mathrm{ch}}=8$, and 32 simultaneous optical connections. Input port--frequency pairs are drawn uniformly without replacement, and output ports are drawn from a rank-weighted distribution (exponent 0.8) that produces unequal output contention. Unavailable output channels are generated as fragmented masks with a common unavailable fraction. All coupling configurations in \cref{fig:global-pinned,fig:connectivity-complexity} are evaluated on the same connection requests and channel-availability masks (seeds 0--29). Supplementary Table~S1 lists the protocol used for every figure.

The converter-bank study uses an explicit $8\times8$ five-stage Benes network with four frequency channels and 16 simultaneous connections. Each $2\times2$ switching element can select its bar or cross state independently for each frequency channel, and ideal frequency converters can be inserted at selected stage boundaries. Each lightpath is allowed at most one frequency conversion. This model is used to compare separate spatial and frequency-conversion stages with the restricted joint space--frequency switching function introduced in Sec.~\ref{sec:spatial-factorization-boundary}.

Assignments are computed with an exact mixed-integer linear program (HiGHS), and only solutions with proven optimality are retained. Full details of the traffic generation, channel-availability masks, converter placement, optimization procedure, and statistical analysis are provided in Supplementary Notes~1 and~3.

\subsection{TFLN synthetic-frequency coupling model}
\label{sec:methods-tfln-converter}

The synthetic-frequency stage is modeled as an electro-optically modulated thin-film lithium niobate (TFLN) resonator with modes separated by one free spectral range (FSR). RF tones applied at integer multiples of the FSR coherently couple modes separated by the corresponding number of frequency spacings. Retaining coupling up to the third FSR, the mode-coupling matrix is

\begin{equation}
h=
\operatorname{diag}(\delta)
+
\sum_{\ell=1}^{3}
J_{\ell}
\left(
e^{i\phi_{\ell}}T_{\ell}
+
e^{-i\phi_{\ell}}T_{\ell}^{\dagger}
\right),
\label{eq:methods-coherent-hamiltonian}
\end{equation}
where $J_{\ell}$ and $\phi_{\ell}$ are the coupling strength and phase of the $\ell$th RF tone, $\delta$ contains the residual modal detunings, and $T_{\ell}$ couples resonator modes separated by $\ell$ FSRs.

The resonator is treated as a linear coherent multi-frequency scattering element using temporal coupled-mode theory~\citep{fan2003tcmt}. Its stationary scattering matrix is

\begin{equation}
S_{\mathrm{res}}(\Omega)
=I-
\tilde{\kappa}_{\mathrm{ext}}
\left[
\left(
\frac{\tilde{\kappa}}{2}-i\Omega
\right)I
+
i\tilde{h}
\right]^{-1},
\label{eq:methods-scattering}
\end{equation}
where $\tilde{h}=2\pi h$, $\tilde{\kappa}$ is the total photon-decay rate in angular-frequency units, $\tilde{\kappa}_{\mathrm{ext}}$ is the external-coupling contribution, and $\Omega$ is the optical detuning from the nominal operating point. Routing calculations use $\Omega=0$, whereas the detuning sweep in Fig.~\ref{fig:tfln-optimized-topology}(e,f) evaluates the same fixed RF configuration as a function of $\Omega$.

The coupling and decay-rate scales are taken from the reported TFLN resonator of Ref.~\citep{dinh2024reconfigurable}, with $J_{1}^{\mathrm{rep}}=0.34$~GHz, $J_{2}^{\mathrm{rep}}=0.24$~GHz, $J_{3}^{\mathrm{rep}}=0.21$~GHz, and $\kappa\equiv2\gamma=1.26$~GHz. For each active-tone set, $J_{\ell}$ and $\phi_{\ell}$ are optimized within bounded ranges using calibration switching instances and are subsequently kept fixed during evaluation. The principal calculation assumes $\eta_{\mathrm{ext}}=\kappa_{\mathrm{ext}}/\kappa=1$, corresponding to the optimistic limit of zero intrinsic resonator loss. Eight frequency channels are embedded in a 16-mode resonator model so that scattering outside the useful switching band is retained.

For an input mode $m$, the delivered power in output mode $n$ is

\begin{equation}
T_{nm}
=
10^{-L_{\mathrm{syn}}/10}
\left|
\left[S_{\mathrm{res}}(0)\right]_{nm}
\right|^{2},
\label{eq:methods-target-power}
\end{equation}
where $L_{\mathrm{syn}}=1$~dB is the additional resonator-plane path loss applied to the complete $T_{nm}$ matrix. Both diagonal and off-diagonal terms of $T_{nm}$ are retained in the routing model. A mode transfer is accepted when
\begin{equation*}
10\log_{10}(T_{nm})\geq T_{\min,\mathrm{dB}},
\end{equation*}
with $T_{\min,\mathrm{dB}}=-6$~dB used as the reference power threshold. The ideal lossless bypass sets the diagonal transmission to unity for frequency-preserving paths while retaining the driven off-diagonal response. It serves only as an architectural upper bound and does not represent a physical device.

The physical switching model uses $N=8$, $N_{\mathrm{ch}}=8$, and $N_{\mathrm{req}}=32$ in a five-stage Benes network, with one synthetic-frequency resonator plane between the second and third stages. A fraction $\rho_{\mathrm{link}}=0.25$ of the frequency channels on each internal link is unavailable. The reference configuration uses $N_{\mathrm{res}}=8$, the architectural assignment limit $n_{\mathrm{sim}}=4$, and one independent RF-control group ($N_{\mathrm{drive}}=1$). Full details of the RF optimization, calibration and evaluation sets, finite-mode convergence, simultaneous-transfer assumptions, and shared-control compatibility are given in Supplementary Notes~2--4.

\subsection{Routing and performance evaluation}
\label{sec:methods-routing-evaluation}

Spatial routes and output frequency modes are optimized jointly. The primary objective is to maximize the number of established optical connections. Among solutions establishing the same number of connections, the configuration with the largest total delivered target-mode power is selected.

The connection-blocking fraction is defined as

\begin{equation}
B_{\mathrm{block}}
=
1-\frac{N_{\mathrm{est}}}{N_{\mathrm{req}}},
\label{eq:methods-blocking}
\end{equation}
where $N_{\mathrm{req}}$ and $N_{\mathrm{est}}$ are the numbers of requested and established connections, respectively.

For the physical TFLN model, each candidate transfer is weighted by the corresponding delivered power $T_{nm}$ obtained from the resonator scattering matrix. In addition to connection blocking, we evaluate the delivered target-mode power, conversion loss for off-diagonal transfers, target-to-strongest in-band non-target mode isolation, and out-of-band scattering. These quantities characterize the optical response of the synthetic-frequency stage and are not intended as receiver-level metrics such as bit-error rate or data throughput.

The same traffic instances, channel-availability patterns, and candidate spatial routes are used for all compared configurations. Detailed definitions of the power and isolation metrics, optimization procedure, RF-control constraints, and statistical evaluation are provided in Supplementary Notes~2 and~3.

\begin{acknowledgments}
This research received no external funding.
\end{acknowledgments}

\section*{Generative AI Use Statement}

Generative AI tools (OpenAI ChatGPT and Anthropic Claude) were used to assist with code development and language editing. The author reviewed and verified all AI-assisted content and takes full responsibility for this work.

\section*{Conflict of Interest}
The author declares no conflict of interest.

\section*{Data Availability Statement}
The source code, calculated scattering matrices, and seed-level data supporting this work are available from the author upon reasonable request.

\section*{Supplemental Material}
Additional supporting information can be found in the separate Supporting Information document.

\bibliography{references}

\end{document}


\title{Supporting Information for: Requirements for Synthetic-Frequency Switching in Integrated Photonic Fabrics}
\author{Jorge Parra}
\email{jorge.parra@uv.es}
\affiliation{Institute of Materials Science (ICMUV), University of Valencia, C/ Catedr\'atico Jos\'e Beltr\'an 2, 46980 Paterna, Valencia, Spain}
\date{November 20, 2026}

\begin{abstract}
This document contains the numerical protocols and notation, analytical derivations, extended numerical and physical methods, and complementary results supporting the main manuscript.
\end{abstract}

\keywords{Photonic switching | Synthetic-frequency photonics | Frequency-mode coupling | Thin-film lithium niobate | Optical circuit switching}
\maketitle

\setlength{\parindent}{0pt}
\setlength{\parskip}{0.45\baselineskip}

\section*{Notation and numerical protocols}
\label{sec:si-protocols}

Table~\ref{tab:si-protocols} lists the numerical protocol behind every figure that reports connection blocking, and Table~\ref{tab:si-notation} summarizes the notation used in the main text and in this document. All idealized calculations in main-text Figs.~\MainFigGlobalPinned{} and~\MainFigConnectivityComplexity{} use the same request generator and the same channel-availability masks for a given seed, so the values in the two figures refer to identical switching instances.

\begin{table*}[t]
\centering
\caption{
Numerical protocols. ``Requests'' gives the number of simultaneous connection requests $N_{\mathrm{req}}$. Input port--frequency pairs are always sampled uniformly without replacement and output ports from the rank-weighted distribution of Eq.~\eqref{eq:si-output-demand}. Unavailability masks use a fragmentation parameter of 0.5 unless stated otherwise. Calibration and evaluation seeds are disjoint whenever a parameter is selected from data.
}
\label{tab:si-protocols}
\small
\begin{tabular}{@{}p{2.4cm}p{3.6cm}p{2.7cm}p{3.1cm}p{\dimexpr\textwidth-11.8cm-8\tabcolsep\relax}@{}}
\hline
\raggedright Figure & \raggedright Model & \raggedright $N$, $N_{\mathrm{ch}}$, requests & \raggedright Unavailability & \raggedright Seeds and selection \tabularnewline
\hline\hline
\raggedright Main 2(a) & \raggedright Idealized assignment & \raggedright 8, 8, 32 & \raggedright Output channels, 0--75\% & \raggedright Evaluation 0--29 \tabularnewline
\raggedright Main 2(b) & \raggedright Idealized assignment & \raggedright 8, 8, 8--64 & \raggedright Output channels, 50\% & \raggedright Evaluation 0--29 \tabularnewline
\raggedright Main 3(a) & \raggedright Idealized assignment & \raggedright 8, \{4, 8, 16\}, $0.5NN_{\mathrm{ch}}$ & \raggedright Output channels, 50\% & \raggedright Evaluation 0--29 \tabularnewline
\raggedright Main 3(b,c) & \raggedright Idealized assignment & \raggedright 8, 8, 32 & \raggedright Output channels, 50\% & \raggedright Evaluation 0--29 \tabularnewline
\raggedright Fig.~\ref{fig:supp_ideal_additional}(a) & \raggedright Idealized assignment & \raggedright 8, 8, 32 & \raggedright Output channels, 50\% & \raggedright Fixed tone pair selected on seeds 0--9; evaluation 10--29 \tabularnewline
\raggedright Fig.~\ref{fig:supp_ideal_additional}(b) & \raggedright Idealized assignment & \raggedright 8, \{4, 8, 16\}, $\{0.5,0.75\}NN_{\mathrm{ch}}$ & \raggedright Output channels, 50\%; fragmentation 0, 0.5, 1 & \raggedright Loads selected in a three-seed pilot (seeds 0--2) among 0.5, 0.75, 1.0; evaluation 0--29 \tabularnewline
\raggedright Main 5(b), Fig.~\ref{fig:supp_converter_planes} & \raggedright Frequency-selective Benes, ideal converters, $\le1$ conversion per lightpath & \raggedright 8, 4, 16 & \raggedright Internal links, 25\% & \raggedright Converter-plane order selected on seeds 30--44; evaluation 0--29 \tabularnewline
\raggedright Main 6(g), Fig.~\ref{fig:supp_nsim} & \raggedright Benes MZI network with TCMT resonator plane ($N_{\mathrm{res}}=8$, $N_{\mathrm{drive}}=1$, $n_{\mathrm{sim}}=4$) & \raggedright 8, 8, 32 & \raggedright Internal links, 25\% & \raggedright RF amplitudes and phases calibrated on seeds 30--44, fixed for evaluation on 0--29 \tabularnewline
\hline
\end{tabular}
\end{table*}

\begin{table*}[t]
\centering
\caption{Notation.}
\label{tab:si-notation}
\small
\begin{tabular}{@{}p{2.6cm}p{\dimexpr0.5\textwidth-2.6cm-2\tabcolsep-0.75em\relax}@{\hspace{1.5em}}p{2.6cm}p{\dimexpr0.5\textwidth-2.6cm-2\tabcolsep-0.75em\relax}@{}}
\hline
Symbol & \raggedright Meaning & Symbol & \raggedright Meaning \tabularnewline
\hline
$N$ & \raggedright Number of physical input/output ports & $J_\ell$, $\phi_\ell$ & \raggedright Coupling strength and phase of the RF tone at $\ell$ FSRs \tabularnewline
$N_{\mathrm{ch}}$ & \raggedright Frequency channels per port & $\kappa$, $\kappa_{\mathrm{ext}}$ & \raggedright Total and external photon-decay rates ($\kappa\equiv2\gamma$) \tabularnewline
$N_{\mathrm{req}}$, $N_{\mathrm{est}}$ & \raggedright Requested and established connections & $\eta_{\mathrm{ext}}$ & \raggedright External-coupling fraction $\kappa_{\mathrm{ext}}/\kappa$ \tabularnewline
$B_{\mathrm{block}}$ & \raggedright Connection blocking, $1-N_{\mathrm{est}}/N_{\mathrm{req}}$ & $\Omega$ & \raggedright Optical detuning from the operating point \tabularnewline
$\mathcal{I}_{\mathrm{RF}}$ & \raggedright Set of active RF-tone orders (in FSR units) & $T_{nm}$ & \raggedright Power delivered from input mode $m$ to output mode $n$ \tabularnewline
$U_\ell$ & \raggedright Share of established connections using separation $\ell$ & $T_{\min,\mathrm{dB}}$ & \raggedright Acceptance threshold on $T_{nm}$ \tabularnewline
$F_{\mathrm{rec}}$ & \raggedright Normalized blocking reduction, Eq.~\eqref{eq:si-frec} & $L_{\mathrm{syn}}$ & \raggedright Resonator-plane path loss \tabularnewline
$\Pi_i$, $U_\nu$ & \raggedright Frequency-independent spatial stage; frequency-only transformation & $N_{\mathrm{res}}$, $N_{\mathrm{drive}}$ & \raggedright Number of resonators and of independent RF-control groups \tabularnewline
$\mathbf{S}_{nm}^{(s,r)}$ & \raggedright $2\times2$ spatial block of a joint element, input mode $m$ to output mode $n$ & $n_{\mathrm{sim}}$ & \raggedright Maximum simultaneous transfers assigned to one resonator \tabularnewline
$N_{\mathrm{sw}}$, $R_{\mathrm{sw}}$ & \raggedright Benes element count; relative reduction in element count & $\rho_{\mathrm{link}}$ & \raggedright Unavailable fraction of channels on each internal link \tabularnewline
$L_{\mathrm{B}}$, $\mathcal{F}_P$, $B_P$ & \raggedright Benes stages; feasible assignments and blocking with $P$ converter banks &  &  \tabularnewline
\hline
\end{tabular}
\end{table*}

\section{Supplementary Note 1 --- Analytical results}
\label{sec:si-note1}

We consider a multiset $\mathcal{R}$ of unit-capacity optical connections between $N$ physical input and output ports, each supporting $N_{\mathrm{ch}}$ orthogonal frequency channels. For a physical port $p$, let $N_{\mathrm{conn}}(p)$ denote the number of connections incident on that port, and define $N_{\mathrm{conn}}^{\max} = \max_{p} N_{\mathrm{conn}}(p)$. Within one switching configuration, each frequency channel at every physical input, output, and internal link has unit capacity. In the ideal model, each established connection is assigned a spatial route and an allowed input--output frequency pair. In the physical model, this ideal frequency assignment is replaced by the resonator scattering response described in Supplementary Note~2.

\subsection{Multiplexing and synthetic-frequency coupling with freely selectable channels}

\textbf{Corollary S1 (Multiplexing--synthetic-coupling equivalence for freely selectable channels).}\label{thm:si-flexible-state-equivalence}
If every connection may use any of the $N_{\mathrm{ch}}$ frequency channels at both physical ports, all channels have equal unit capacity, there are no shared frequency-coupling capacity constraints, and the multiplexed fabric can independently switch each frequency channel, then

\begin{equation}
N_{\mathrm{config}}^{\mathrm{mux}}
=
N_{\mathrm{config}}^{\mathrm{syn}}
=
\left\lceil
\frac{N_{\mathrm{conn}}^{\max}}{N_{\mathrm{ch}}}
\right\rceil .
\label{eq:si-flexible-slot-equivalence}
\end{equation}

Here, $N_{\mathrm{conn}}^{\max}$ is the maximum number of optical connections associated with any physical input or output port.

\textit{Proof.}
Construct the bipartite multigraph $G_{\mathcal{R}}$ whose left and right vertices represent the physical input and output ports and whose edge multiset is $\mathcal{R}$. Each switching configuration provides only $N_{\mathrm{ch}}$ unit-capacity channels at each physical port. Therefore, both architectures require at least

\begin{equation*}
\left\lceil
\frac{N_{\mathrm{conn}}^{\max}}{N_{\mathrm{ch}}}
\right\rceil
\end{equation*}

configurations.

By K\"onig's line-coloring theorem, the bipartite multigraph $G_{\mathcal{R}}$ admits a proper edge coloring using exactly $N_{\mathrm{conn}}^{\max}$ colors. These colors can be partitioned into groups containing at most $N_{\mathrm{ch}}$ colors each. Each group is assigned to one switching configuration, with its colors mapped injectively to distinct optical frequency channels. Since all edges of a given color form a matching, no channel-resolved input or output port is used more than once within that configuration. For a rearrangeable spatial fabric, each partial matching can be completed to a full permutation when required.

This construction establishes all requested connections in

\begin{equation*}
\left\lceil
\frac{N_{\mathrm{conn}}^{\max}}{N_{\mathrm{ch}}}
\right\rceil
\end{equation*}

configurations using ordinary frequency-preserving multiplexed switching. A photonic switching fabric with ideal synthetic-frequency coupling can reproduce every such assignment and is subject to the same port-capacity lower bound. The two architectures therefore require the same minimum number of switching configurations. \hfill$\square$

This result specializes the standard edge-coloring construction used in optical switch scheduling to the present switching model~\citep{berry2005transceiver,aggarwal2003edgecoloring,deneve2025edgecoloring}. It shows that, when frequency channels can be selected freely, ideal synthetic-frequency coupling changes the available frequency-channel mappings but does not increase the number of simultaneous orthogonal channels carried by a physical port.

\subsection{Frequency conversion and spatial routing}

\textbf{Proposition S2 (A frequency-only transformation does not create frequency-dependent spatial routing).}\label{prop:si-space-state-factorization}
Let $\Pi_1,\Pi_2\in\mathbb{C}^{N\times N}$ denote frequency-independent spatial permutations, let $U_\nu\in\mathbb{C}^{N_{\mathrm{ch}}\times N_{\mathrm{ch}}}$ denote a frequency-only transformation, and let $I_N$ and $I_\nu$ be the identity operators in the spatial and frequency-channel spaces, respectively. Then

\begin{equation}
(\Pi_2\otimes I_\nu)
(I_N\otimes U_\nu)
(\Pi_1\otimes I_\nu)
=
(\Pi_2\Pi_1)\otimes U_\nu .
\label{eq:si-space-state-factorization}
\end{equation}

Consequently, the set of reachable physical output ports is independent of the input frequency channel. If all three operators are permutations, the transformation reduces to $(i,m)\mapsto \bigl(\pi(i),\sigma(m)\bigr)$, so every frequency channel experiences the same spatial permutation.

\textit{Proof.}
Using the mixed-product property of the tensor product,

\begin{equation}
(\Pi_2\otimes I_\nu)
(I_N\otimes U_\nu)
(\Pi_1\otimes I_\nu)
=
(\Pi_2 I_N\otimes I_\nu U_\nu)
(\Pi_1\otimes I_\nu),
\label{eq:si-space-state-factorization-intermediate}
\end{equation}

and therefore

\begin{equation}
(\Pi_2\otimes I_\nu)
(I_N\otimes U_\nu)
(\Pi_1\otimes I_\nu)
=
(\Pi_2\Pi_1)\otimes U_\nu .
\label{eq:si-space-state-factorization-proof}
\end{equation}

For an input basis state $|i,m\rangle$, the amplitude at output state $|j,n\rangle$ is proportional to $(\Pi_2\Pi_1)_{ji}(U_\nu)_{nm}$. The spatial factor is independent of the input frequency index $m$, proving that a separate frequency-only transformation cannot introduce frequency-dependent spatial connectivity. \hfill$\square$

\subsection{Programmable-element comparison}

We next compare different architectures capable of implementing independent spatial permutations for the available frequency channels. The target family is

\begin{equation}
\mathcal{T}_{N,N_{\mathrm{ch}}}
=
\left\{
\operatorname{diag}
(\Pi_1,\ldots,\Pi_{N_{\mathrm{ch}}})
:
\Pi_m\in\mathfrak{S}_N
\right\},
\label{eq:si-element-target-family}
\end{equation}

with $|\mathcal{T}_{N,N_{\mathrm{ch}}}| = (N!)^{N_{\mathrm{ch}}}$. For $N=2^k$, a binary $N\times N$ Benes network contains $L(N)=2\log_2(N)-1$ stages and

\begin{equation}
N_{\mathrm{sw}}(N)
=
\frac{N}{2}L(N)
\label{eq:si-benes-count}
\end{equation}

programmable $2\times2$ switching sites.

An architecture composed of $N_{\mathrm{ch}}$ independent frequency-preserving Benes planes supports the complete target family $\mathcal{T}_{N,N_{\mathrm{ch}}}$. We denote this architecture by S.

Consider instead an architecture SC$_0$ composed of one frequency-independent Benes network followed or preceded by a frequency-only conversion stage. Although this architecture may reach different frequency channels, Proposition~S2 restricts its frequency-preserving spatial transformations to a common spatial permutation for all channels. Its coverage of the target family is therefore

\begin{equation}
\frac{
|\mathcal{T}_{N,N_{\mathrm{ch}}}\cap\mathrm{SC}_0|
}{
|\mathcal{T}_{N,N_{\mathrm{ch}}}|
}
=
\frac{N!}{(N!)^{N_{\mathrm{ch}}}}
=
\frac{1}{(N!)^{N_{\mathrm{ch}}-1}}.
\label{eq:si-sc0-coverage}
\end{equation}

Thus, the lower nominal number of spatial switching sites in SC$_0$ does not represent a reduction relative to an equivalent switching function.

A joint space--frequency architecture, denoted SC, instead allows each Benes site at stage $s$ and row $r$ to implement

\begin{equation}
\mathbf{S}_{\mathrm{sf}}^{(s,r)}
=
\sum_{m,n=1}^{N_{\mathrm{ch}}}
\mathbf{S}_{nm}^{(s,r)}
\otimes
|n\rangle\langle m|,
\qquad
\mathbf{S}_{nm}^{(s,r)}
\in
\mathbb{C}^{2\times2},
\label{eq:si-state-conditioned-site}
\end{equation}

with
 $\mathbf{S}_{\mathrm{sf}}^{(s,r)} \neq \Pi^{(s,r)} \otimes U_\nu^{(s,r)}$ in the general case.

Functional equivalence with the independent frequency-preserving planes requires, at minimum, the independently programmable diagonal subset

\begin{equation}
\mathbf{S}_{\mathrm{sf,diag}}^{(s,r)}
=
\sum_{m=1}^{N_{\mathrm{ch}}}
B^{(s,r)}
\bigl(b_{s,r,m}\bigr)
\otimes
|m\rangle\langle m|,
\label{eq:si-state-conditioned-diagonal-subset}
\end{equation}

where $B^{(s,r)}(b_{s,r,m})$ selects the local bar or cross state for frequency channel $m$. Controlled off-diagonal blocks $\mathbf{S}_{nm}^{(s,r)}$ with $n\neq m$ additionally provide coherent frequency conversion.

Under the convention of counting one ideal nonseparable primitive as one spatial switching site, the joint architecture contains $N_{\mathrm{sw}}(N)$ spatial sites instead of $N_{\mathrm{ch}}N_{\mathrm{sw}}(N)$ sites in the independent-plane architecture. This gives the nominal spatial-site reduction

\begin{equation*}
R_{\mathrm{spatial}}
=
1-\frac{1}{N_{\mathrm{ch}}}.
\end{equation*}

However, this counting does not establish a reduction in physical components or control channels. The joint architecture still requires $N_{\mathrm{ch}}N_{\mathrm{sw}}(N)$ independently programmable frequency-dependent bar/cross decisions, in addition to the controls needed for off-diagonal frequency conversion. If the diagonal response in Eq.~\eqref{eq:si-state-conditioned-diagonal-subset} is implemented using $N_{\mathrm{ch}}$ conventional MZIs per site, the MZI count returns to $N_{\mathrm{ch}}N_{\mathrm{sw}}(N)$.

For $N=8$ and $N_{\mathrm{ch}}=4$, four independent Benes planes require 80 conventional $2\times2$ elements, and the ideal joint architecture contains 20 joint elements. A second conventional implementation demultiplexes the channels and routes the resulting $NN_{\mathrm{ch}}$ port--frequency tributaries through one larger Benes network, with each output restricted to the corresponding multiplexer input. For $N$ and $N_{\mathrm{ch}}$ that are powers of two, it requires

\begin{equation}
N_{\mathrm{sw}}^{\mathrm{spatialized}}
=
N_{\mathrm{sw}}(NN_{\mathrm{ch}})
=
N_{\mathrm{ch}}N_{\mathrm{sw}}(N)
+
NN_{\mathrm{ch}}\log_2(N_{\mathrm{ch}})
\label{eq:si-spatialized-benes-count}
\end{equation}

elements, that is, 144 in the same example. The additional term appears because the larger network supports arbitrary permutations among all tributaries, while the required transformation is block diagonal. The diagonal subset of the joint architecture and the independent planes require the same number of independently programmable bar/cross decisions, so any hardware saving depends on how efficiently a physical joint element realizes these decisions.

\subsection{Spatially conditioned frequency conversion}

We finally compare the ideal joint architecture with a conventional frequency-selective spatial network supplemented by independently controlled frequency converters.

\textbf{Proposition S3 (Nested converter-plane hierarchy).}
Consider a frequency-selective Benes network with $L_B=2\log_2(N)-1$ spatial stages. A converter plane placed after a spatial stage contains one independently controlled and bypassable frequency converter on every spatial branch. Let $\mathcal{P}$ denote the set of installed converter planes, and let $\mathcal{F}(\mathcal{P})$ denote the set of feasible connection assignments when each optical path may undergo at most one ideal frequency conversion. If $\mathcal{P}\subseteq\mathcal{Q}$, then

\begin{equation}
\mathcal{F}(\mathcal{P})
\subseteq
\mathcal{F}(\mathcal{Q}).
\label{eq:si-converter-plane-nesting}
\end{equation}

When converter planes are installed after all $L_B$ spatial stages,

\begin{equation}
\mathcal{F}
\bigl(
\{0,\ldots,L_B-1\}
\bigr)
=
\mathcal{F}_{\mathrm{joint}}^{(1)},
\label{eq:si-converter-plane-equivalence}
\end{equation}

where $\mathcal{F}_{\mathrm{joint}}^{(1)}$ denotes the feasible set of the restricted joint-site model in which each optical path may undergo at most one conversion after a selected spatial site.

\textit{Proof.}
Adding converter planes cannot remove a feasible assignment because every added converter can be bypassed. Therefore, if $\mathcal{P}\subseteq\mathcal{Q}$, $\mathcal{F}(\mathcal{P}) \subseteq \mathcal{F}(\mathcal{Q})$. For the second statement, consider any schedule allowed by the restricted joint-site model. Its single conversion event occurs after one of the $L_B$ spatial stages and on the branch selected by the corresponding spatial route. A separate architecture containing converter planes after every stage provides an independently controlled converter at exactly that stage boundary and on that branch. Replacing the joint-site conversion with this branch converter preserves both the spatial route and every link--frequency occupancy. Conversely, every conversion performed by a branch converter is also permitted by the restricted joint-site model. The two feasible sets are therefore identical. \hfill$\square$

The corresponding blocking fractions obey

\begin{equation}
B_{\mathrm{joint}}^{(1)}
=
B_{L_B}
\leq
B_{L_B-1}
\leq
\cdots
\leq
B_1
\leq
B_0.
\label{eq:si-converter-plane-blocking}
\end{equation}

This equivalence applies only to the restricted single-conversion routing model. A general nonseparable $2N_{\mathrm{ch}}\times2N_{\mathrm{ch}}$ joint element may implement coherent space--frequency transformations that cannot be decomposed into a frequency-selective spatial switch followed by independent branch converters.

With $P$ converter planes, the separate architecture contains $N_{\mathrm{sw}}(N)$ frequency-selective elements and $NP$ branch converters, and each lightpath traverses $L_B$ switching elements and up to $P$ converters. A comparison of cost or insertion loss with the joint architecture would require the properties of a physical joint element, which has not been demonstrated, so no numerical weights are assigned here.

\subsection{Converter-plane numerical comparison}

The numerical comparison in main-text Fig.~\MainFigSpatialElementScaling{}(b) uses an explicit $8\times8$ five-stage Benes topology with $N_{\mathrm{ch}}=4$ and 16 simultaneously requested optical connections. Each request specifies an input port, an output port, and a fixed input frequency. Input port--frequency pairs are sampled uniformly without replacement, while output ports follow the same rank-weighted distribution used in the other Benes calculations.

Every spatial switching element is frequency selective, allowing its bar or cross state to be chosen independently for each frequency channel. The input and output frequencies of a conversion are optimized jointly with the spatial route, and each optical path may undergo at most one ideal frequency conversion.

For every internal link, 25\% of the frequency channels are initially unavailable in one circular contiguous block. A fragmentation parameter of 0.5 exchanges unavailable and available channel positions while preserving the total unavailable fraction. Input and output links remain fully available.

Starting from $P=0$, converter planes are added sequentially. At each step, the next stage-output plane is selected to maximize the mean number of admitted connections over calibration seeds 30--44. This produces the one-indexed nested order $\ConverterPlaneCalibrationOrder{}$, which is then kept fixed for evaluation on seeds 0--29. The connection assignment is solved exactly using a mixed-integer linear program that maximizes the number of established connections subject to one assignment per request, unit occupancy of each link--frequency channel, and compatible frequency-dependent bar/cross states at every spatial switching site.

Connection blocking decreases from \ConverterPlaneZeroBlocking{} without frequency conversion to \ConverterPlaneOneBlocking{} after adding the first converter plane. The minimum number of converter planes that reproduces the admission of the restricted joint-site architecture for every evaluation seed is $P^\star=\ConverterPlanePStar$, for which the blocking fraction is \ConverterPlaneFullBlocking{}. This implementation contains \ConverterPlanePStarBranches{} independently controlled branch converters in addition to the 20 frequency-selective spatial switching sites, giving \ConverterPlanePStarTotalBlocks{} nominal functional elements. The restricted joint abstraction instead contains 20 composite space--frequency sites. The corresponding \JointSiteNominalBlockReduction{} reduction follows only if these heterogeneous functional elements are assigned equal weight and should not be interpreted as a physical component reduction.

The fifth converter plane lies on the output branches and does not further reduce blocking for the present workload because the endpoint frequency channels remain available and the output frequency can be selected freely. This result is specific to the considered traffic and channel-availability patterns. The analytical hierarchy in Eq.~\eqref{eq:si-converter-plane-blocking} remains valid more generally, but the minimum required number of converter planes $P^\star$ must be determined for the relevant operating conditions.

The seed-level results and the converter-plane calibration are available from the author upon request.

\begin{figure}[t]
    \centering
    \includegraphics[width=0.34\textwidth]{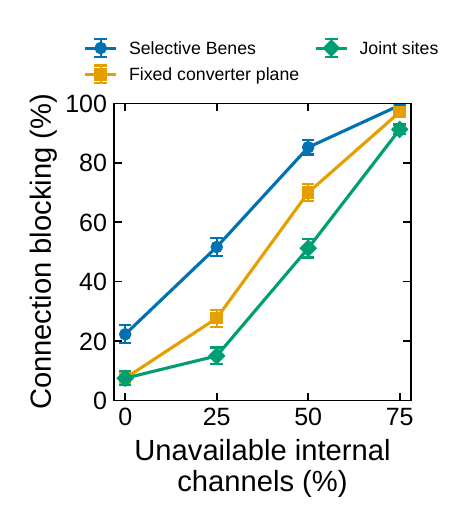}
    \caption{
    Connection blocking as the fraction of unavailable internal frequency channels is varied. The frequency-selective Benes network preserves the carrier frequency, the fixed-plane architecture permits one ideal conversion after the second stage, and the restricted joint-site architecture permits one ideal conversion after any selected spatial site. Points show means over seeds 0--29, with 95\% confidence intervals.
    }
    \label{fig:supp_converter_planes}
    \label{fig:si-joint-site-unavailability}
\end{figure}

An additional channel-unavailability sweep (Fig.~\ref{fig:supp_converter_planes}) compares a frequency-selective Benes network without conversion, the same network with one fixed converter plane after the second stage, and the restricted joint-site architecture allowing one ideal conversion after any selected spatial site. The same connection requests and channel masks are used for all three cases. Conversion loss, crosstalk, RF-control compatibility, and target-mode power thresholds are omitted in this ideal comparison.

\section{Supplementary Note 2 --- Thin-film lithium niobate resonator model}
\label{sec:si-note2}

The synthetic-frequency stage is modeled as an electro-optically modulated thin-film lithium niobate (TFLN) resonator using temporal coupled-mode theory (TCMT). RF modulation at the free spectral range (FSR) and its integer multiples couples resonator modes separated by the corresponding frequency spacings. The model distinguishes the coupling and decay-rate scales reported for the reference device from the architectural assumptions introduced for the switching analysis, and explicitly retains sideband power outside the eight frequency channels used by the switching fabric.

\subsection{TCMT model and shared RF control}

An RF tone at $\ell$ times the FSR directly couples resonator modes separated by $\ell$ frequency spacings. Because the optical field evolves coherently within the resonator, however, repeated coupling can also transfer amplitude over mode separations that are not directly driven. The resulting stationary response therefore depends on the complete multimode coupling matrix together with the external and intrinsic decay rates.

We define the shift operator $T_{\ell}$ such that $(T_{\ell})_{mn}=1$ when $m=n+\ell$. Retaining RF coupling up to the third FSR, the frequency-mode coupling matrix is

\begin{equation}
h_{\mathrm{GHz}}
=
\operatorname{diag}(\boldsymbol{\delta})
+
\sum_{\ell=1}^{3}
J_{\ell}
\left(
e^{i\phi_{\ell}}T_{\ell}
+
e^{-i\phi_{\ell}}T_{\ell}^{\dagger}
\right),
\label{eq:si-literature-coherent}
\end{equation}

where $J_{\ell}$ and $\phi_{\ell}$ are the coupling strength and RF phase associated with the $\ell$th tone, and $\delta_k$ is the residual detuning of resonator mode $k$. The reported simultaneous-fit values

\begin{equation*}
J_1^{\mathrm{rep}}=0.34~\mathrm{GHz},\qquad
J_2^{\mathrm{rep}}=0.24~\mathrm{GHz},\qquad
J_3^{\mathrm{rep}}=0.21~\mathrm{GHz},
\end{equation*}

together with the reported photon-decay scale $2\gamma=1.26~\mathrm{GHz}$, are taken from the reference TFLN synthetic-frequency resonator~\citep{dinh2024reconfigurable}. We use the convention $\kappa\equiv2\gamma=1.26~\mathrm{GHz}$ for the total photon-decay rate. These values define the coupling and decay-rate scales of the present model, but they are not taken as a complete description of the measured device or of its bus-waveguide coupling condition.

For the TCMT equations, angular-frequency quantities are defined as

\begin{equation*}
\tilde{h}=2\pi h_{\mathrm{GHz}},
\qquad
\tilde{\kappa}_x=2\pi\kappa_{x,\mathrm{GHz}},
\end{equation*}

with $x\in\{\mathrm{ext},\mathrm{int}\}$ and $\kappa=\kappa_{\mathrm{ext}}+\kappa_{\mathrm{int}}$. Using the phasor convention $e^{-i\omega t}$, the intracavity amplitudes $\mathbf{a}$ satisfy

\begin{equation}
\dot{\mathbf{a}}
=
\left(
-i\tilde{h}
-
\frac{\tilde{\kappa}}{2}I
\right)\mathbf{a}
+
\sqrt{\tilde{\kappa}_{\mathrm{ext}}}\,\mathbf{s}_{+},
\label{eq:si-tcmt-state}
\end{equation}

with the input--output relation

\begin{equation*}
\mathbf{s}_{-}
=
\mathbf{s}_{+}
-
\sqrt{\tilde{\kappa}_{\mathrm{ext}}}\,\mathbf{a}.
\end{equation*}

At the nominal operating point, the corresponding complex scattering matrix is

\begin{equation}
S_{\mathrm{res}}(0)
=
I
-
\tilde{\kappa}_{\mathrm{ext}}
\left(
\frac{\tilde{\kappa}}{2}I
+
i\tilde{h}
\right)^{-1}.
\label{eq:si-tcmt-scattering}
\end{equation}

For an optical detuning $\Omega$ from this operating point,

\begin{equation}
S_{\mathrm{res}}(\Omega)
=
I
-
\tilde{\kappa}_{\mathrm{ext}}
\left[
\left(
\frac{\tilde{\kappa}}{2}
-
i\Omega
\right)I
+
i\tilde{h}
\right]^{-1}.
\label{eq:si-tcmt-bandwidth}
\end{equation}

The model assumes mode-independent external coupling and intrinsic decay over the retained frequency modes. These quantities were not independently measured for every mode in the reference device.

Routing calculations use $S_{\mathrm{res}}(0)$. The optical-detuning calculations retain the same optimized RF amplitudes and phases while evaluating Eq.~\eqref{eq:si-tcmt-bandwidth}. The mode-resolved spectra and out-of-band leakage of main-text Fig.~\MainFigTFLN{}(e,f) span $|\Omega/2\pi|\leq5\kappa$ and are plotted against $\Omega/\tilde{\kappa}=(\Omega/2\pi)/\kappa$. The converted-path analysis of Fig.~\ref{fig:supp_fixed_control} spans

\begin{equation*}
-2\kappa
\leq
\frac{\Omega}{2\pi}
\leq
2\kappa.
\end{equation*}

These sweeps describe the stationary response of a fixed RF configuration and should not be interpreted as a modulation-bandwidth or data-transmission model.

Eight central resonator modes define the useful switching band. To account for scattering outside this band, the main calculation embeds these channels in a 16-mode resonator model. Models containing 8, 16, 24, and 32 modes are used to verify convergence of the calculated out-of-band power. The main calculation assumes

\begin{equation*}
\eta_{\mathrm{ext}}
=
\frac{\kappa_{\mathrm{ext}}}{\kappa}
=
1,
\end{equation*}

corresponding to the optimistic limit $\kappa_{\mathrm{int}}=0$, and applies an additional resonator-plane path loss $L_{\mathrm{syn}}=1$~dB to the complete response, including both diagonal and off-diagonal terms.

For an input mode $m$ and output mode $n$, the delivered power is

\begin{equation}
T_{nm}
=
10^{-L_{\mathrm{syn}}/10}
\left|
[S_{\mathrm{res}}(0)]_{nm}
\right|^2.
\label{eq:si-delivered-target-power}
\end{equation}

The matrix $T$ therefore contains both the physical diagonal response, corresponding to frequency-preserving transmission, and the off-diagonal response associated with frequency conversion.

A transfer is accepted by the routing model when $T_{nm,\mathrm{dB}} = 10\log_{10}(T_{nm}) \geq T_{\min,\mathrm{dB}}$, including the case $n=m$. The reference value is $T_{\min,\mathrm{dB}}=-6.0~\mathrm{dB}$, which corresponds to 25\% delivered target-mode power. This value is used as a screening criterion rather than as a receiver sensitivity or device specification, and the full threshold dependence is reported in the main text. An on-chip lithium-niobate frequency shifter has previously demonstrated transfer efficiencies of approximately 90\%~\citep{hu2021frequency}, illustrating that the chosen threshold is not intended as a fundamental efficiency limit.

The per-separation medians in main-text Fig.~\MainFigTFLN{}(d) are taken over the ordered off-diagonal entries of $T$, whereas the converted-path values in Supplementary Note~4 are taken over the established connections that undergo frequency conversion.

For an individual mode transfer $m\rightarrow n$, the target-to-strongest in-band non-target mode isolation is

\begin{equation}
I_{nm}
=
10\log_{10}
\left(
\frac{
T_{nm}
}{
\displaystyle
\max_{\substack{k\neq n\\k\in\mathcal{B}}}
T_{km}
}
\right),
\label{eq:si-mode-isolation}
\end{equation}

where $\mathcal{B}$ denotes the eight-mode switching band. The converted-path detuning analysis retains the paths selected at $\Omega=0$.

The coupling plane is further characterized by the number of physical resonators $N_{\mathrm{res}}$, the assumed maximum number of simultaneous transfers assigned to one resonator $n_{\mathrm{sim}}$, and the number of independently controlled RF groups $N_{\mathrm{drive}}\leq N_{\mathrm{res}}$. Resonators belonging to the same RF-control group are optically distinct but share the same nominal RF amplitudes and phases and therefore the same scattering matrix $S_{\mathrm{res}}$. The main calculation uses $N_{\mathrm{res}}=8, \qquad n_{\mathrm{sim}}=4, \qquad N_{\mathrm{drive}}=1$. The parameter $n_{\mathrm{sim}}$ is an architectural assignment constraint and does not follow from the TCMT equations or from a measured physical capacity of the resonator.

\subsection{Model parameters and RF optimization}

Table~\ref{tab:tfln_parameters} summarizes the physical parameters, architectural assumptions, and numerical search ranges used in the TFLN calculation. Literature-derived quantities are kept separate from assumptions introduced specifically for the switching model.

\begin{table}[t]
\centering
\caption{
Physical parameters, architectural assumptions, and numerical ranges used in the TFLN synthetic-frequency model.
}
\label{tab:tfln_parameters}
\small
\begin{tabular}{@{}lll@{}}
\hline
Quantity & Value or range & Role \\
\hline
$J_1^{\mathrm{rep}},J_2^{\mathrm{rep}},J_3^{\mathrm{rep}}$
&
$0.34,\,0.24,\,0.21$ GHz
&
Reported coupling scales \\

$J_{\ell}$
&
$0\leq J_{\ell}\leq2J_{\ell}^{\mathrm{rep}}$
&
RF optimization bounds \\

$\kappa\equiv2\gamma$
&
$1.26$ GHz
&
Reported photon-decay scale \\

$\phi_1,\phi_2,\phi_3$
&
$[-\pi,\pi]$
&
RF phase bounds \\

Switching/resonator modes
&
8 embedded in 16
&
\shortstack[l]{Useful band and OOB\\scattering} \\

$\eta_{\mathrm{ext}}$
&
1
&
Optimistic $\kappa_{\mathrm{int}}=0$ limit \\

$L_{\mathrm{syn}}$
&
1 dB
&
\shortstack[l]{Additional resonator-plane path loss\\applied to the full response} \\

$T_{\min,\mathrm{dB}}$
&
$-6$ dB
&
Reference transfer threshold \\

$N_{\mathrm{res}},n_{\mathrm{sim}},N_{\mathrm{drive}}$
&
$8,\,4,\,1$
&
Architecture and RF control \\

$\lambda_{\mathrm{leak}}$
&
0.25
&
\shortstack[l]{Off-target penalty in\\RF calibration} \\

$\Omega/2\pi$
&
$\pm5\kappa$; $\pm2\kappa$
&
Detuning sweeps (spectra; converted paths) \\
\hline
\end{tabular}
\end{table}

For each active-tone configuration, the RF coupling amplitudes and phases are calibrated using bounded deterministic multistart optimization. The three initial conditions are the reported coupling amplitudes with zero RF phases and two seed-controlled perturbations. Each start is optimized using L-BFGS-B for at most 70 iterations. Active coupling strengths vary independently over $0\leq J_{\ell}\leq2J_{\ell}^{\mathrm{rep}}$, while their phases vary over $-\pi\leq\phi_{\ell}\leq\pi$. Inactive RF tones remain exactly zero.

For the 1+2-FSR configuration, the optimized parameters are $(J_1,J_2) = (\NestedDOneTwoJOne,\NestedDOneTwoJTwo)~\mathrm{GHz}$, with $(\phi_1,\phi_2) = (\NestedDOneTwoPhiOne,\NestedDOneTwoPhiTwo)~\mathrm{rad}$. When the third RF tone is also permitted, the optimization gives

\begin{equation*}
(J_1,J_2,J_3)
=
(\NestedRichJOne,\NestedRichJTwo,\NestedRichJThree)~\mathrm{GHz},
\end{equation*}

with

\begin{equation*}
(\phi_1,\phi_2,\phi_3)
=
(\NestedRichPhiOne,\NestedRichPhiTwo,\NestedRichPhiThree)~\mathrm{rad}.
\end{equation*}

These are bounded local-search solutions and are not claimed to be globally optimal.

RF settings are calibrated using switching instances generated with seeds 30--44 and are then kept fixed during evaluation on seeds 0--29. This separation prevents the RF parameters from being reoptimized for the evaluation instances.

For RF-control group $g$, let $\mathcal{T}_g$ denote the multiset of requested input--output mode transfers in the calibration cases. With $S=S_{\mathrm{res}}(0,\boldsymbol{\theta})$, where $\boldsymbol{\theta}$ collects the RF amplitudes and phases, the calibration maximizes

\begin{equation}
\mathcal{J}_g(\boldsymbol{\theta})
=
\frac{1}{|\mathcal{T}_g|}
\sum_{(m,n)\in\mathcal{T}_g}
\left[
|S_{nm}|^2
-
0.25
\left(
\sum_{k\in\mathcal{M}_{\mathrm{phys}}}
|S_{km}|^2
-
|S_{nm}|^2
\right)
\right],
\label{eq:si-rf-objective}
\end{equation}

where $\mathcal{M}_{\mathrm{phys}}$ denotes the retained resonator modes. The objective therefore maximizes the mean target-mode power while penalizing total power scattered into non-target modes.

Residual mode detunings are initialized at zero and constrained to $|\delta_k|\leq10^{-9}~\mathrm{GHz}$, so independent tuning of every retained optical resonance is not introduced as an additional hardware degree of freedom.

\subsection{RF configurations and scattering matrices}

The stationary scattering matrix supplied to the routing optimizer is the fixed $\Omega=0$ response defined in Eq.~\eqref{eq:si-tcmt-scattering}.

The principal physical calculation uses 16 resonator modes, with the central eight assigned to the switching fabric. Larger retained-mode models are used only for the finite-mode convergence analysis of the out-of-band response.

The calibrated RF setting is shared by all resonators belonging to a given RF-control group. Consequently, individual mode-to-mode transfers cannot generally be optimized independently when several connections use the same control setting. This shared-control constraint is examined explicitly in Supplementary Note~4 by comparing independently optimized transfers with their best jointly supported response.

The calibrated RF configurations and the calculated scattering and power-transfer matrices are available from the author upon request.

\section{Supplementary Note 3 --- Numerical implementation}
\label{sec:si-note3}

This section describes the numerical implementation used for the idealized switching calculations and for the physical Benes-network model. It includes the generation of connection requests and channel-availability patterns, the routing constraints, the optimization objectives, and the reference configurations used throughout the main text.

\subsection{Idealized switching under frequency-channel constraints}
\label{sec:si-ideal-assignment}

The idealized calculations underlying Figs.~\MainFigGlobalPinned{} and~\MainFigConnectivityComplexity{} of the main text consider $N$ input and output ports, each supporting $N_{\mathrm{ch}}$ frequency channels. The reference case uses $N=8,\qquad N_{\mathrm{ch}}=8,\qquad N_{\mathrm{req}}=32$ simultaneous unit-capacity optical connections.

For each random instance, the $N_{\mathrm{req}}$ input port--frequency pairs $(i,m)$ are sampled uniformly without replacement from the $NN_{\mathrm{ch}}$ available combinations. Each requested connection therefore occupies a distinct input channel. Output ports are sampled from a nonuniform distribution intended to introduce unequal traffic loading without reproducing a specific measured network trace. A random permutation assigns a rank $r_j\in\{1,\ldots,N\}$ to each output port, with sampling probability

\begin{equation}
p_j
=
\frac{r_j^{-0.8}}
{\displaystyle\sum_{k=1}^{N}r_k^{-0.8}}.
\label{eq:si-output-demand}
\end{equation}

For each point of the unavailable-output-channel sweep, the unavailable fraction is fixed globally to one of $0,\;0.25,\;0.50,\;0.625,\;0.75$. A distinct frequency-channel-availability mask is then generated independently for each output port using this common fraction. For a given unavailable fraction, the corresponding channels initially form a circular contiguous block with a uniformly sampled starting position. The pattern is subsequently fragmented by exchanging unavailable and available channel positions while preserving the total unavailable fraction. The number of exchanges is

\begin{equation*}
\operatorname{round}
\left[
0.5\,
\min
\left(
n_{\mathrm{unavailable}},
N_{\mathrm{ch}}-n_{\mathrm{unavailable}}
\right)
\right],
\end{equation*}

corresponding to the fragmentation parameter of 0.5 used in the main calculations. Independent random-number streams are used for connection generation and channel-availability masks.

Each switching instance is evaluated using the same connection requests and availability pattern for all coupling configurations. A connection with fixed input mode $m$ may be assigned an output mode $n$ according to

\begin{equation}
\begin{aligned}
\text{frequency preserving:}\qquad
& n=m,\\
\text{nearest neighbor:}\qquad
& n=m
\;\text{or}\;
|n-m|=1,\\
\text{first three spacings:}\qquad
& n=m
\;\text{or}\;
|n-m|\leq3,\\
\text{unrestricted:}\qquad
& n\in\{1,\ldots,N_{\mathrm{ch}}\}.
\end{aligned}
\label{eq:si-ideal-coupling-cases}
\end{equation}

More generally, restricted synthetic-frequency coupling is specified by the active RF-tone-order set $\mathcal{I}_{\mathrm{RF}}$ introduced in the main text.

Spatial assignment and output frequency are optimized jointly rather than selected sequentially for each connection. The optimization maximizes the number of established connections subject to output-channel availability and unit occupancy of every input and output frequency channel. The resulting mixed-integer linear programs are solved using the SciPy MILP interface to the HiGHS solver, and only solutions returned with proven-optimal status are retained.

For each optimized solution, the utilization of frequency separation $\ell$ is obtained by counting established connections satisfying $|n-m|=\ell$ and dividing by the total number of established connections. The categories $\ell=0,1,2,3,$ and $\ell>3$ form a complete partition of the admitted connections.

All paired comparisons use seeds 0--29. Plotted values are seed means, and the reported 95\% confidence intervals are calculated using the normal approximation,

\begin{equation*}
\bar{x}
\pm
1.96\frac{s}{\sqrt{30}}.
\end{equation*}

Because the same seeds are used across architectures, differences between configurations are evaluated on paired switching instances.

The normalized-load sweep in main-text Fig.~\MainFigGlobalPinned{}(b) uses
\begin{equation*}
N_{\mathrm{req}}\in\{8,16,24,32,40,48,56,64\},
\end{equation*}
corresponding to
\begin{equation*}
\frac{N_{\mathrm{req}}}{NN_{\mathrm{ch}}}
\in
\{0.125,0.25,0.375,0.5,0.625,0.75,0.875,1.0\}
\end{equation*}
at fixed 50\% output-channel unavailability.

To compare coupling sets across workloads, the blocking reduction of a configuration with blocking $B$ is normalized on each instance to the reduction obtained with unrestricted coupling,
\begin{equation}
F_{\mathrm{rec}}=\frac{B_{0}-B}{B_{0}-B_{\mathrm{full}}},
\label{eq:si-frec}
\end{equation}
where $B_0$ and $B_{\mathrm{full}}$ are the frequency-preserving and unrestricted-coupling blocking of the same instance. Instances with $B_0=B_{\mathrm{full}}$ are excluded from the average. Reported values of $F_{\mathrm{rec}}$ are means over instances, which can differ slightly from the ratio of mean blocking values.

The channel-count comparison of main-text Fig.~\MainFigConnectivityComplexity{}(a) uses the same generator with $N=8$, $N_{\mathrm{ch}}\in\{4,8,16\}$, $N_{\mathrm{req}}=0.5NN_{\mathrm{ch}}$, 50\% output-channel unavailability, and a fragmentation parameter of 0.5. For $N_{\mathrm{ch}}=8$, these instances coincide with those of main-text Fig.~\MainFigGlobalPinned{}. The broader comparison of Fig.~\ref{fig:supp_ideal_additional}(b) averages over two normalized loads and three fragmentation parameters (0, 0.5, and 1). The two loads were selected, for each $N_{\mathrm{ch}}$, among 0.5, 0.75, and 1.0 as those with the largest difference between frequency-preserving and unrestricted blocking in a three-seed pilot (seeds 0--2). This selection only concentrates the evaluation on informative regimes and does not use the coupling configurations being compared. For the fixed-pair comparison of Fig.~\ref{fig:supp_ideal_additional}(a), the fixed two-tone configuration is selected on seeds 0--9 and all configurations are evaluated on seeds 10--29.

\subsection{Benes routing and traffic generation}

The physical switching calculations use an explicit $8\times8$ five-stage Benes network. Each internal spatial link supports eight frequency channels, and one synthetic-frequency resonator plane is placed between the second and third MZI stages.

A candidate lightpath specifies the spatial route through the Benes network, the bar/cross state required at each traversed MZI, the frequency channel occupied on every internal link, the resonator used at the synthetic-frequency plane, and the selected input--output frequency transfer. Within a switching configuration, incompatible states of the same physical MZI and repeated occupation of the same link--frequency channel are forbidden.

Frequency-preserving paths are treated using the diagonal transmission $T_{mm}$ of the corresponding scattering matrix. Frequency-converting paths use the appropriate off-diagonal term $T_{nm}$.

Each evaluation instance contains 32 optical connection requests with distinct input port--frequency pairs. Output ports and fixed input frequencies are generated deterministically from the random seed using the same traffic model described above. A fraction $\rho_{\mathrm{link}}=0.25$ of the frequency channels on each internal link is marked unavailable without removing the underlying spatial link. All compared physical configurations receive identical connection requests, candidate spatial routes, and channel-availability masks. Evaluation uses seeds 0--29.

The converter-bank calculations of main-text Fig.~\MainFigSpatialElementScaling{}(b) use the related $8\times8$ five-stage Benes model with $N_{\mathrm{ch}}=4$ and 16 simultaneous connection requests. The complete converter-placement procedure, internal-channel masks, calibration seeds, and restricted single-conversion model are given in Supplementary Note~1.

\subsection{Routing objective and reference configurations}

Spatial routes and output modes are optimized jointly to maximize the number of established connections, and the largest total delivered target-mode power breaks ties. Blocking is defined as in main-text Eq.~(13). RF settings are calibrated on seeds 30--44 and held fixed for evaluation on seeds 0--29 (Supplementary Note~2). Because resonators in the same RF-control group share one scattering matrix, the router must select transfers that are compatible with that common response.

Three reference configurations separate the effects of resonator transmission and frequency conversion. The RF-off reference uses the diagonal response of the undriven resonator with the same 1-dB path penalty as the driven case. The ideal lossless bypass keeps the driven off-diagonal response but assigns unit transmission to frequency-preserving paths, which bounds the benefit available if the stage could be bypassed whenever no conversion is required. Finally, the unrestricted-coupling reference of the idealized model allows any channel transfer without loss or selectivity constraints. None of the metrics include receiver filtering, detector noise, or bit-error rate, so they describe connection admission and mode-resolved power transfer within the fabric rather than link performance.

\section{Supplementary Note 4 --- Complementary results}
\label{sec:si-note4}

This section reports additional calculations that support the architectural and device-level results presented in the main text. They cover alternative RF-tone selections in the idealized model, the two- and three-tone TFLN responses, the simultaneous-transfer limit, finite-mode convergence, the detuning response of the converted paths, and the compatibility of transfers under a shared RF configuration.

\subsection{Additional idealized switching comparisons}

\begin{figure}[t]
    \centering
    \includegraphics[width=0.88\textwidth]{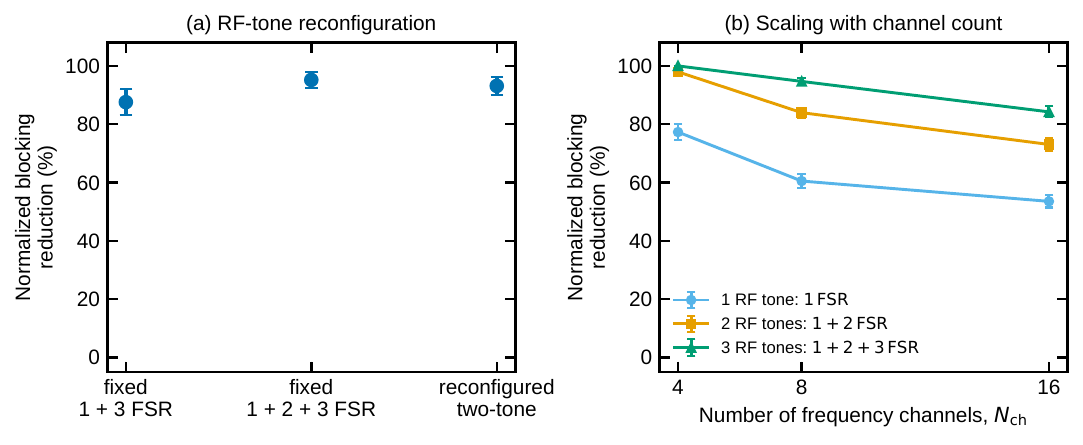}
    \caption{
    Additional idealized switching results. (a) Normalized blocking reduction for fixed 1+3-FSR and 1+2+3-FSR coupling, together with an oracle benchmark that selects the best two-tone configuration for each switching instance. (b) Blocking reduction versus number of frequency channels for coupling at one FSR, one and two FSRs, and the first three FSRs. Values are normalized to the unrestricted-coupling reference, and error bars indicate 95\% confidence intervals over seeds 0--29.
    }
    \label{fig:supp_ideal_additional}
    \label{fig:si-connectivity-secondary}
\end{figure}

The idealized switching model was first used to examine whether the choice of directly coupled frequency separations can be improved by reconfiguring the active RF tones for each switching instance. For the $N_{\mathrm{ch}}=8$ workload, selecting the best two-tone configuration independently for each instance gives only a modest improvement over the fixed 1+3-FSR configuration [Fig.~\ref{fig:supp_ideal_additional}(a)]. This reconfigured case represents an oracle benchmark in which the preferred coupling pattern is known for each switching instance, and it is not intended as an implemented online RF-control strategy.

We also examine how a fixed coupling range behaves as the number of frequency channels increases [Fig.~\ref{fig:supp_ideal_additional}(b)]. Coupling at one FSR, at one and two FSRs, and at the first three FSRs is compared for $N_{\mathrm{ch}}=4$, 8, and 16. The blocking reduction is normalized to that obtained with unrestricted inter-mode coupling for the same switching instances. As $N_{\mathrm{ch}}$ increases, a fixed number of directly coupled separations spans a progressively smaller fraction of the available frequency-channel space and recovers a smaller fraction of the unrestricted-coupling benefit. This result complements the mode-separation analysis in Fig.~\MainFigConnectivityComplexity{} of the main text and reinforces that the useful coupling range depends on the number of channels and on their availability constraints.

\subsection{Comparison of the 1+2-FSR and 1+2+3-FSR resonator responses}

The main physical reference uses RF tones at one and two times the FSR. To isolate the effect of adding a third directly driven mode separation, we compare this configuration with an independently calibrated 1+2+3-FSR drive. Both RF settings are obtained using the bounded multistart procedure described in Supplementary Note~2 and are evaluated using the same optical connections, channel masks, resonator model, number of resonators, simultaneous-transfer limit, RF-control grouping, path loss, and evaluation seeds.

The optimized 1+2-FSR response uses $(J_1,J_2) = (\NestedDOneTwoJOne,\NestedDOneTwoJTwo)~\mathrm{GHz}$, whereas the three-tone optimization gives

\begin{equation*}
(J_1,J_2,J_3)
=
(\NestedRichJOne,\NestedRichJTwo,\NestedRichJThree)~\mathrm{GHz}.
\end{equation*}

Because the optimized $J_3$ is very small, the two mode-to-mode power-transfer matrices are nearly identical at $\Omega=0$.

At the reference threshold $T_{\min,\mathrm{dB}}=-6.0$~dB, the ideal lossless bypass gives \NestedBypassBoundBlocking{} connection blocking. The driven resonator gives \NestedDOneTwoBlocking{}, compared with \NestedRFOffBlocking{} for the RF-off resonator. The through-channel terms of the driven response remain below the acceptance threshold, while seven ordered off-diagonal transfers satisfy the threshold. Among established converted paths in the fabric-level evaluation, the 1+2-FSR configuration gives a mean converted-path power of \NestedDOneTwoConvertedPower{}, a median converted-path loss of \NestedDOneTwoConvertedLoss{}~dB, and a median mode isolation of \NestedDOneTwoModeIsolation{}~dB. Adding the third RF tone does not change the fabric-level blocking.

\begin{figure}[t]
    \centering
    \includegraphics[width=0.36\textwidth]{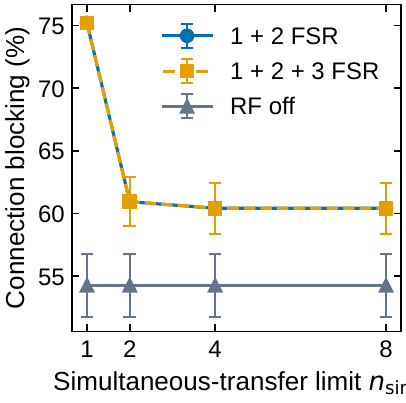}
    \caption{
    Connection blocking versus the assumed simultaneous-transfer limit $n_{\mathrm{sim}}$. The 1+2-FSR and 1+2+3-FSR configurations use their independently calibrated RF settings at $T_{\min,\mathrm{dB}}=-6$~dB. The RF-off reference uses the same connection requests, channel masks, spatial routes, and resonator count. Error bars indicate 95\% confidence intervals over seeds 0--29.
    }
    \label{fig:supp_nsim}
    \label{fig:si-nested-simultaneous-transfer-limit}
\end{figure}

We next test the sensitivity of this result to the assumed maximum number of simultaneous optical transfers assigned to one resonator. For the 1+2-FSR response, blocking decreases from \NestedDOneTwoMOneBlocking{} at $n_{\mathrm{sim}}=1$ to \NestedDOneTwoMTwoBlocking{} at $n_{\mathrm{sim}}=2$ and \NestedDOneTwoMFourBlocking{} at $n_{\mathrm{sim}}=4$ [Fig.~\ref{fig:supp_nsim}]. Increasing the limit further to $n_{\mathrm{sim}}=8$ produces no additional reduction. The same saturation is observed for the 1+2+3-FSR configuration. Thus, $n_{\mathrm{sim}}=4$ is not the limiting constraint for the workload used in the main calculation, although $n_{\mathrm{sim}}$ remains an architectural assumption rather than a measured resonator capacity.

\subsection{Finite-mode convergence and detuning response}

\begin{figure}[t]
    \centering
    \includegraphics[width=0.56\textwidth]{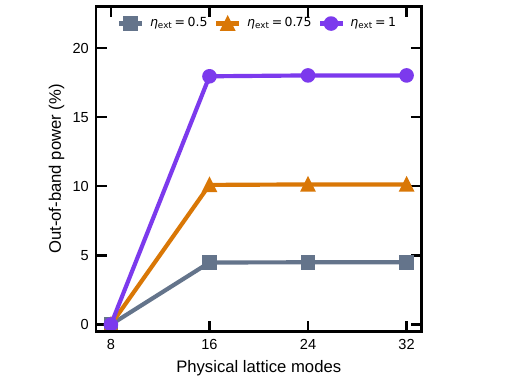}
    \caption{
    Finite-mode convergence of the calculated out-of-band scattering. The eight frequency channels used by the switching fabric are embedded in resonator models containing 8, 16, 24, and 32 modes. The main calculation uses 16 retained modes.
    }
    \label{fig:supp_mode_convergence}
    \label{fig:si-out-of-band}
\end{figure}

Because electro-optic modulation can scatter optical power beyond the eight modes used by the switching fabric, the retained resonator-mode count must be large enough to represent out-of-band sidebands. Fig.~\ref{fig:supp_mode_convergence} compares resonator models containing 8, 16, 24, and 32 modes. The eight-mode model necessarily excludes out-of-band channels and therefore underestimates this contribution. The 16-mode model used in the main calculation captures the additional scattered power and remains consistent with the larger 24- and 32-mode calculations over the tested external-coupling conditions.

\begin{figure}[t]
    \centering
    \subfloat[]{%
    \includegraphics[width=0.36\textwidth]{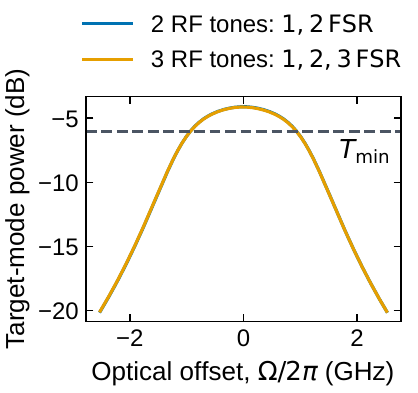}}\hfill
    \subfloat[]{%
    \includegraphics[width=0.36\textwidth]{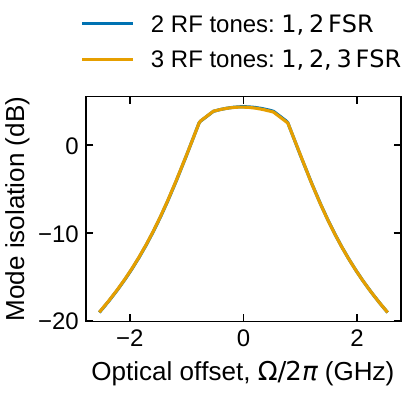}}
    \caption{
    Fixed-control optical-detuning response of the converted paths selected at $\Omega=0$. (a) Mean delivered target-mode power and (b) target-to-strongest in-band non-target mode isolation for the calibrated 1+2-FSR and 1+2+3-FSR configurations. RF controls and selected transfers remain fixed throughout the sweep.
    }
    \label{fig:supp_fixed_control}
    \label{fig:si-detuning-response}
\end{figure}

The fixed-control optical response is further characterized by varying the optical detuning while keeping the calibrated RF amplitudes and phases unchanged. Fig.~\ref{fig:supp_fixed_control}(a) shows the mean target-mode power of the converted paths selected at $\Omega=0$, and Fig.~\ref{fig:supp_fixed_control}(b) shows the corresponding mode isolation. These curves quantify the spectral variation of the stationary scattering matrix around the operating point.

\subsection{Compatibility under a shared RF configuration}

The main physical model assumes that several optical connections may share the same RF-control setting. This requirement differs from optimizing each mode transfer independently because a single multimode scattering matrix must simultaneously support all requested transfers assigned to the same RF-control group.

To quantify this constraint, we consider all 56 ordered off-diagonal transfers among the eight useful frequency modes. Each transfer is first optimized individually within the RF parameter range of Supplementary Note~2. We then form every pair of individually feasible transfers and optimize a common RF configuration to maximize the weaker of the two delivered target-mode powers. Each shared-control optimization uses 60 deterministic starting points and includes the same 1-dB post-scattering path penalty used in the physical switching calculations.

\begin{figure}[t]
    \centering
    \includegraphics[width=0.98\textwidth]{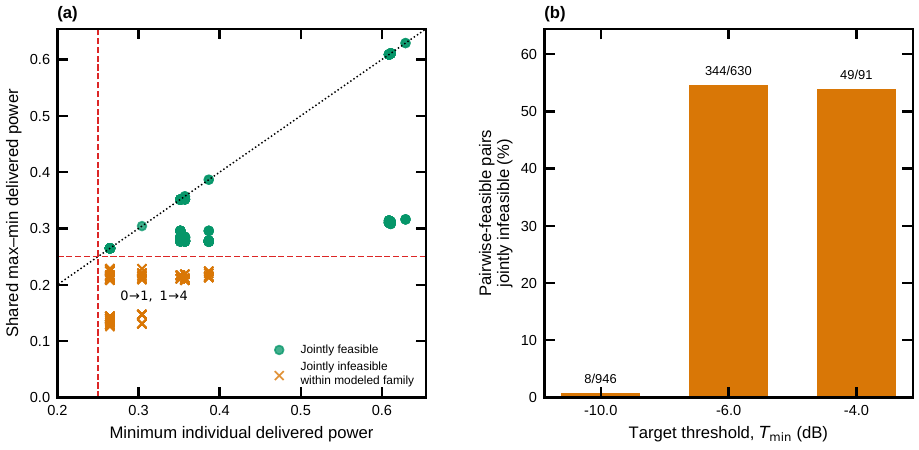}
    \caption{
    Compatibility of mode transfers under a shared RF configuration. (a) Smaller individually optimized delivered power versus the best shared-RF max--min delivered power for pairs of mode transfers. (b) Fraction of individually feasible transfer pairs that cannot be simultaneously supported above the specified target-mode threshold within the explored RF parameter range.
    }
    \label{fig:supp_shared_rf}
    \label{fig:si-joint-compatibility}
\end{figure}

Fig.~\ref{fig:supp_shared_rf}(a) compares the smaller of the two independently optimized target-mode powers with the best common-RF max--min value. Points below the diagonal identify transfer pairs whose performance degrades when both transfers must be supported by one RF setting. The effect becomes particularly important near the operating threshold.

At target-mode thresholds of $-10$, $-6$, and $-4$~dB, the analysis contains \CompatibilityPairsTen{}, \CompatibilityPairsTwentyFive{}, and \CompatibilityPairsForty{} pairs, respectively, for which both transfers are individually feasible. Of these, \CompatibilityIncompatibleTen{}, \CompatibilityIncompatibleTwentyFive{}, and \CompatibilityIncompatibleForty{} pairs cannot be simultaneously supported above the corresponding threshold within the explored RF parameter range. These values correspond to \CompatibilityFractionTen{}, \CompatibilityFractionTwentyFive{}, and \CompatibilityFractionForty{} of the individually feasible pairs [Fig.~\ref{fig:supp_shared_rf}(b)].

The large incompatible fraction around the $-6$~dB operating point shows that individual mode-conversion performance alone can substantially overestimate the set of transfers available under shared RF control. Conversely, failure to identify a common solution within the bounded numerical search does not establish that the corresponding pair is fundamentally impossible, since it only indicates that no compatible RF configuration was found within the modeled parameter family and search range.

\clearpage
\bibliography{references}